\documentclass[final,5p,times,twocolumn]{elsarticle}

\usepackage{epsfig}
\usepackage{graphicx}
\usepackage{subcaption}
\usepackage{booktabs}
\usepackage{epstopdf}
\usepackage{bm}
\usepackage{amsmath} 
\usepackage[figuresright]{rotating}
\usepackage{diagbox}
\usepackage{amssymb}
\usepackage{enumitem}
\usepackage{xcolor}
\usepackage{siunitx}
\usepackage{amssymb}
\usepackage{amsmath}
\usepackage{amsthm}

\usepackage{lineno}

\journal{arXiv}

\begin{document}

\begin{frontmatter}



\title{Evaluation of the performance of an analytical-numerical coupled method for droplet impacts on soft material surfaces}

\author[a]{Hao Hao\corref{co}}
\cortext[co]{Corresponding author}
\ead{hh3117@ic.ac.uk}
\author[a]{Antonis Sergis}
\author[a]{Alex M. K. P. Taylor}
\author[a]{Yannis Hardalupas}
\author[a]{Maria N. Charalambides}

\affiliation[a]{organization={Department of Mechanical Engineering, Imperial College London},
            city={London},
            postcode={SW7 2AZ}, 
            country={United Kingdom}}

\begin{abstract}
Impacts between droplets and solid surfaces can commonly cause erosion problem in Engineering applications,
including aircraft surface erosion, wind blade leading-edge erosion and steam turbine blade erosion. In practice, the impacted solid surfaces have varied material softness, ranging from stiff metallic coatings to soft materials. An analytical-numerical coupled model (ANCM) for simulating droplet impacts on surfaces, and corresponding material analysis, has been developed in the literature. However, the analytical impact pressure solution of the ANCM model has been derived assuming rigid solid surface. In the current study, we investigate the performance of the ANCM model for droplet impacts on soft materials made of urethane gel phantom, by comparing the ANCM computations to lab-based experiments and numerical simulations based on Smoothed Particle Hydrodynamics (SPH). Parametric studies explore the applicability limit of the ANCM model for droplet impacts on very soft materials at low Young's modulus. It was found that for materials at Young's modulus of $47,400$ \unit{\pascal} or stiffer, which covers most engineering applications, the developed ANCM model performs as expected for an assumed rigid surface. For softer solid materials, the SPH-modeled liquid interacts with the evolving surface geometry and mitigates impact intensity as deformation occurs. The analytical impact loads estimated by ANCM are independent of surface geometry, and hence provide conserved impact impulse in a non-physical way. Results show a critical value of Young's modulus at $E=10,000$ \unit{\pascal} for the ANCM model, below which the model exhibits overshoot in total contact force and surface deformation, leading to the formation of steep wall craters.
\end{abstract}




\end{frontmatter}



\section{Introduction}
\label{introduction}
Impact of liquid droplets on solid surfaces has been a research topic for decades due to its ubiquity in the environment. Developed technologies have produced a wide range of applications including aqueous jet-cleaning~\cite{Pergamalis}, $3$D printing~\cite{Yu}, combustion~\cite{Panao}, deposition in pharmacology and agriculture~\cite{Bergeron}. In some engineering applications, understanding the material stress field induced by liquid impacts is important for erosion damage. This includes liquid impacts onto stiff materials, such as erosion problems on aircraft~\cite{Gohardani} and wind turbine blades~\cite{Hao,Hao2}, and soft materials, such as skin-injection ~\cite{Tagawa} and soft matter rupture~\cite{Shojima} in biomedical fields.

Early experimental work on single droplet impact to solid materials noticed the relation between the dynamic loadings of liquid impacts and stress waves in the solids~\cite{Field2}. Particularly, a ring-shape erosion pattern has been widely observed in experiments~\cite{Field2, Engel1, Brunton1} and regarded as the main characteristic of damage on the impact surface. However, erosion studies of stiff materials are typically focused on the material surface, while stress distributions inside the material upon liquid droplet impacts are better revealed on soft materials, and hence (relatively) large deformation, with digital image correlation (DIC). Sun \textit{et al.}~\cite{Sun} embedded tracer particles in polydimethylsiloxane (PDMS) gel and successfully viewed the impact induced stresses using a laser illumination and high-speed stress microscopy. More recently, Yokoyama \textit{et al.}~\cite{Yuto2023, Yuto2024} measured a stressed gelatine gel's photoelastic parameters (phase retardation and azimuthal angle) using a polarization camera and constructed the three-dimensional stress field in the gelatine gel using integrated photoelasticity. Nevertheless, the experimental methods for stress field measurements inside solid materials require transparent materials, which greatly limits applicability.

\begin{figure}
    \begin{subfigure}[b]{\textwidth}
        \includegraphics[width=8 cm]{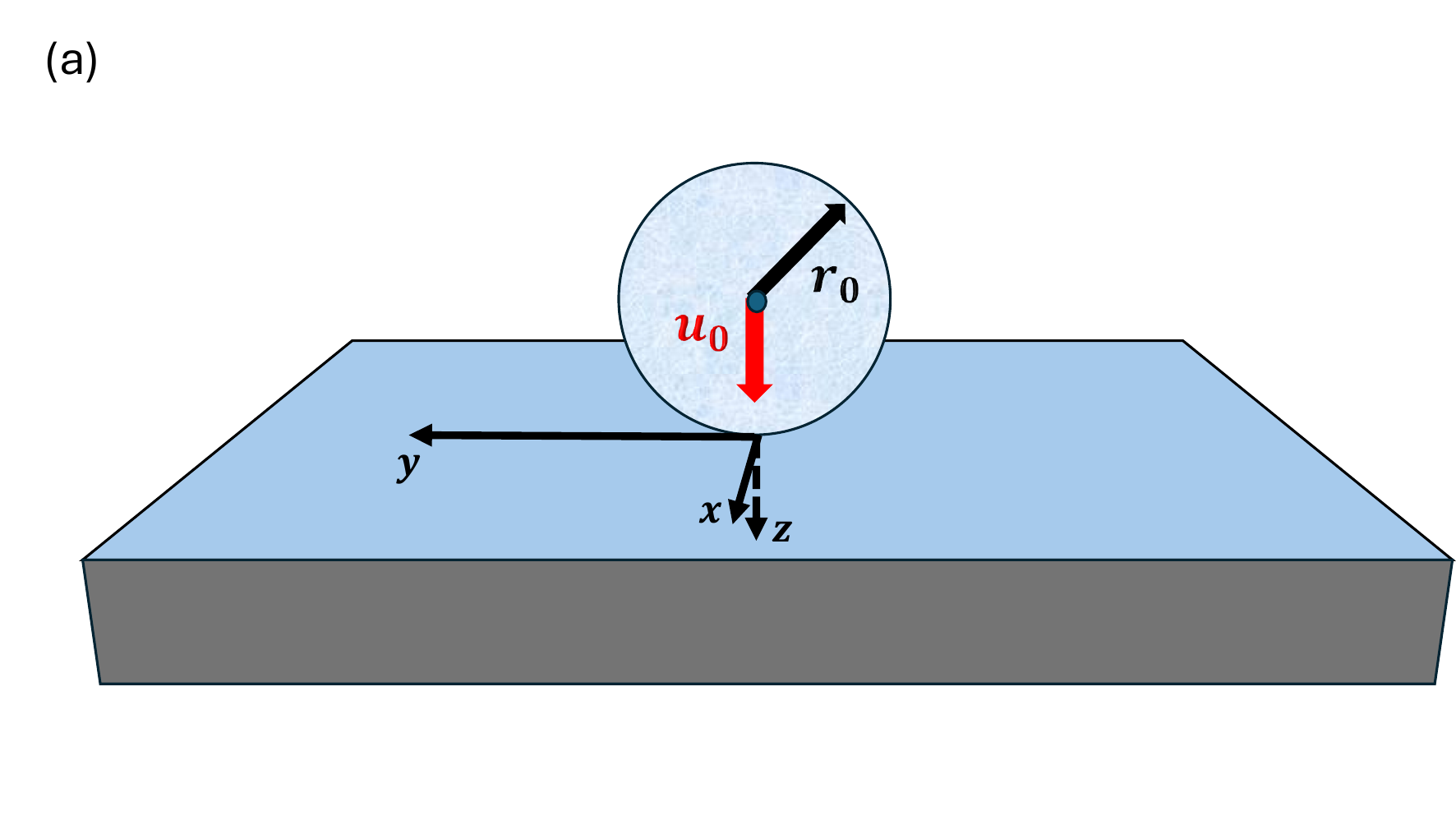}
    \end{subfigure}
    \begin{subfigure}[b]{\textwidth}
        \includegraphics[width=9.5 cm]{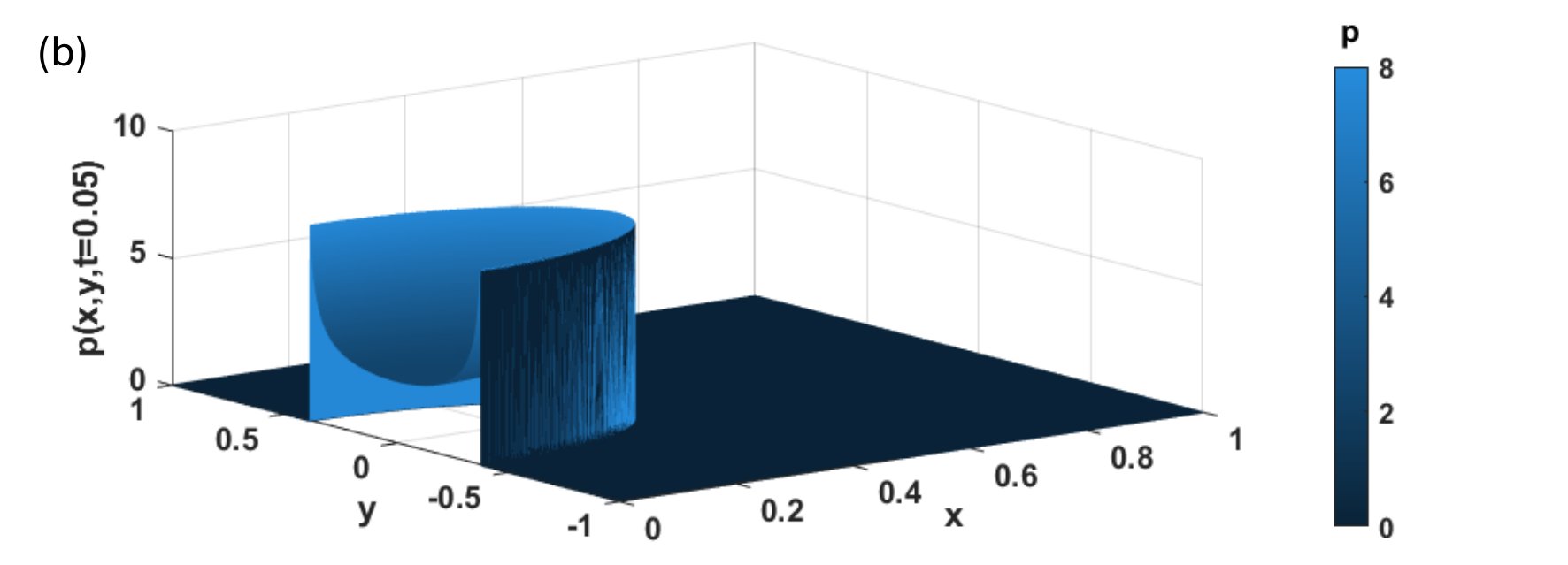}
    \end{subfigure}
    \begin{subfigure}[b]{\textwidth}
        \includegraphics[width=9.5 cm]{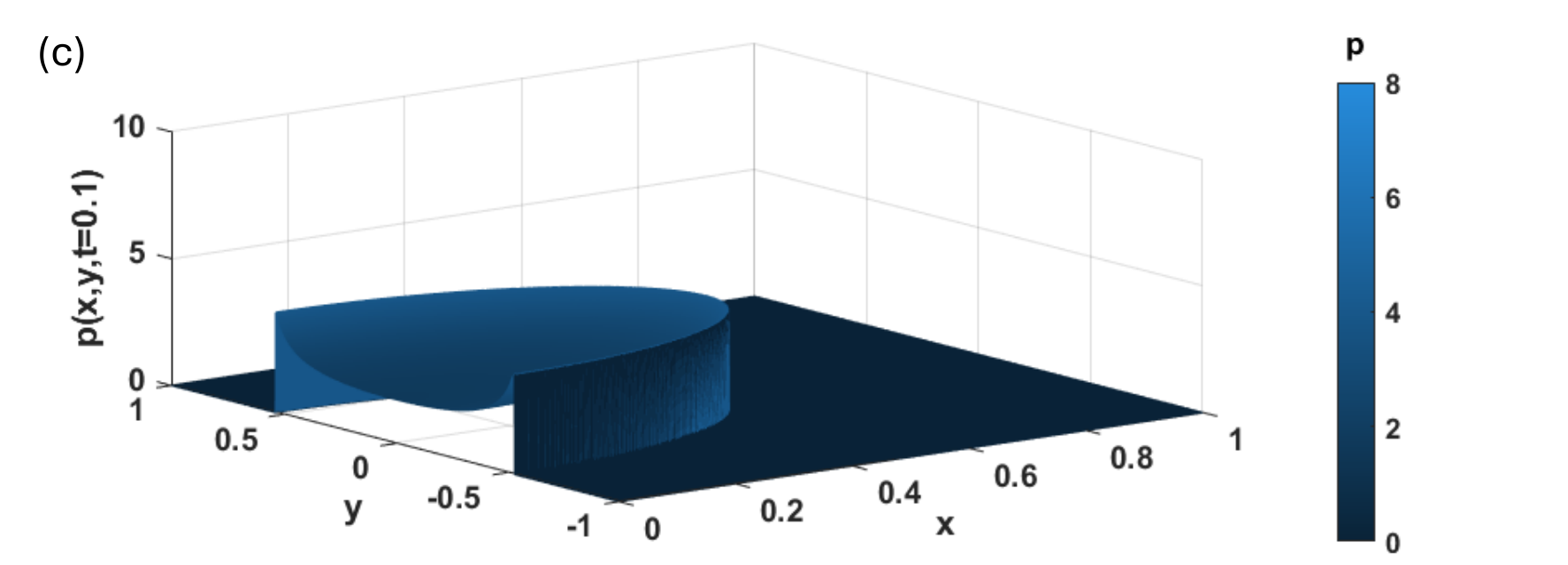}
    \end{subfigure}
    \begin{subfigure}[b]{\textwidth}
        \includegraphics[width=9.5 cm]{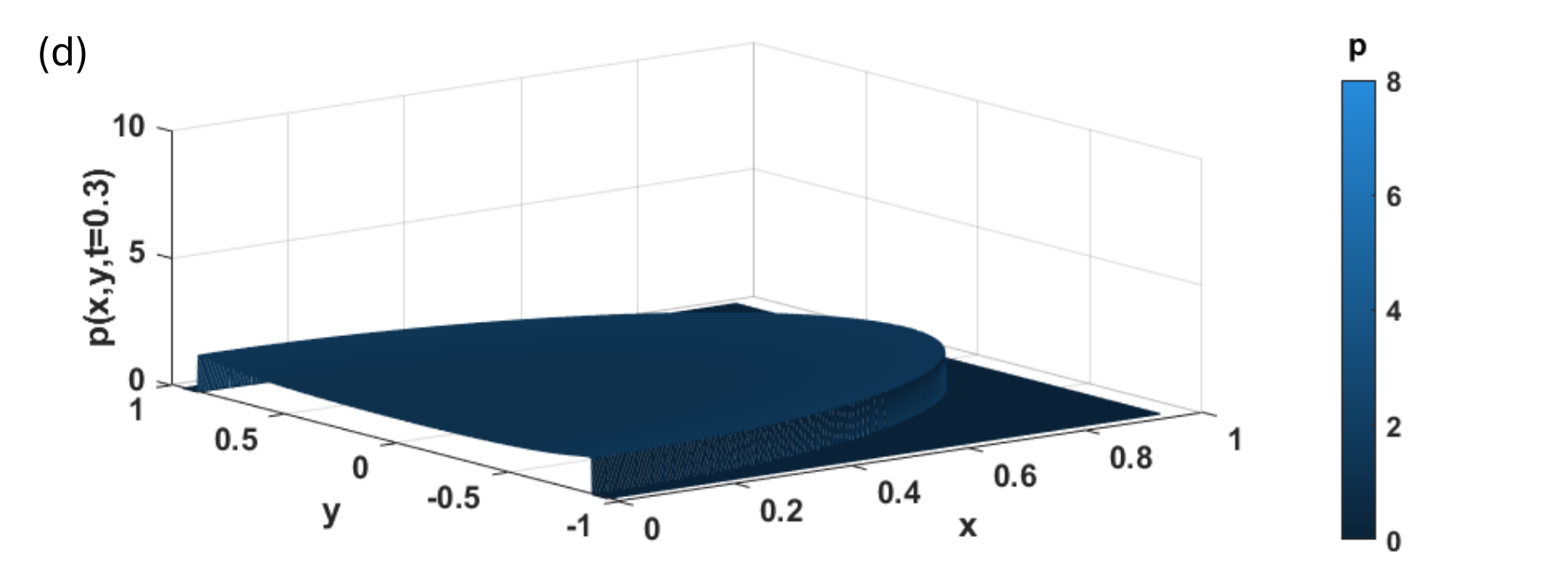}
    \end{subfigure}
    \caption{Initial droplet (and material) configuration in the Cartesian coordinate system (a) and corresponding analytical pressure profiles on the rigid surface upon liquid droplet impact in~\cite{Hao_ANCM}, shown at time $t=0.05$ (b), $t=0.1$ (c) and $t=0.3$ (d). $r_0=1$ and $u_0=1$ denote the radius and impact velocity of the droplet, respectively. Subfigures (b-d) have the same axes and pressure limits. All variables in the figure are presented in their dimensionless forms.}
\label{ring}
\end{figure}

Also, numerical fluid-structure interaction (FSI) models have been developed to explore liquid-solid impacts, which, in principle, are not limited to material choices. Typical FSI models are based on a Finite Element (FE) approach for the solid phase where material deformations, and hence stresses and strains, are discretized and calculated on Lagrangian nodes~\cite{Hrennikoff, Courant}. The general FE approach of material analysis is applicable to both stiff~\cite{Zhou1, XJ} and soft~\cite{Samaras, Bikos} materials. For impacted liquids, FE-based approaches, including Coupled Eulerian-Lagrangian (CEL)~\cite{Keegan1} and Smoothed Particle Hydrodynamics (SPH)~\cite{Leon}, have high computational costs due to large deformations of the liquids~\cite{Verma1} and poor resolutions in capturing the detailed flow features~\cite{Zhou2}. This can be improved in coupled approaches~\cite{Amirzadeh1, Nick2023} which involve a separate Computational Fluid Dynamics (CFD) simulation for the impacted liquids as the dynamic loading conditions on the contact surface of the FE material analysis. The coupled method benefits from better accuracy on the droplet flow dynamics, but requires an extra consideration on the deliberate design of grids at the interface for data exchange. More recently, Hao \textit{et al.}~\citep{Hao_ANCM} developed an analytical-numerical coupled method (ANCM) which incorporated an analytical solution~\citep{Hao_ana} of single liquid droplet impact (Figure \ref{ring}) as a spatio-temporal boundary conditions on the material surface of the FE analysis. The analytical solution replaces the need to numerically simulate the droplet and hence simplifies the contact problem to a single phase FE analysis. Consequently, the developed ANCM requires reduced mesh resolution and provides accurate simulations at a remarkably low computational cost. Nevertheless, we note that all the above-mentioned coupled methods, and most literature on FSI simulations~\cite{Ye}, are one-way coupling algorithms, which assume rigid solid surface geometry with no surface deformation felt by the liquids during impact. In contrast, two-way coupling interchanged information between phases on the contact surface leads to fluid flows that are affected by structural deformation and vice versa~\cite{Richter}. This process is often very complex and requires iterations within timesteps to achieve convergence and desired accuracy~\cite{Ahamed}.

Therefore, the literature review shows that the advantages and potential of the one-way coupled FSI model of ANCM that provides accurate and computationally cheap solutions for engineering applications affected by droplet impacts on solid materials. However, since the analytical solution of droplet impact assumes rigid solid surface, the capabilities of the ANCM model remain unknown for soft material surfaces. The capabilities of the ANCM model to surfaces of softer materials, which have surface deformation, must be evaluated. It is important to develop physical understanding on the subtle changes in material deformation behaviour as stiffness decreases, and hence establish the applicability limits of the rigid surface assumption for soft materials. In this way, a prediction shall be provided for the critical value of material stiffness for the applicability limit of the ANCM model to soft materials, which is the purpose of the present study. Section \ref{methods} introduces the FE numerical models for liquid droplet impact onto solid materials, including a conventional SPH model and the ANCM. Section \ref{results} evaluates the simulation results through comparisons to experimental data, followed by a parametric study of the effect of material properties. Observed trends and differences are discussed, which provides a critical material stiffness for the applicability limit of the ANCM model to soft materials. Finally, Section \ref{conclusion} summarizes the main conclusions.

\section{Methods}
\label{methods}
\subsection{Analytical liquid droplet impact pressure}
\label{2.1}
\begin{figure}
\centering
\includegraphics[width=9.5 cm]{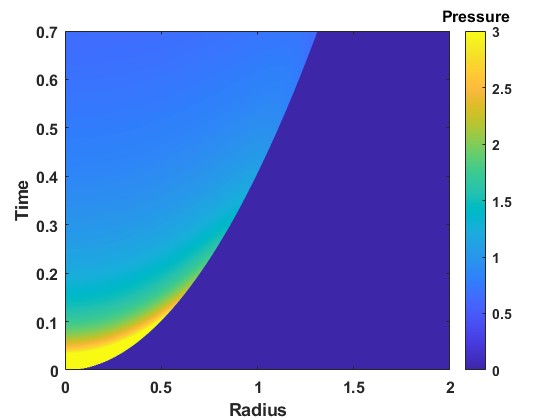}	
\caption{The analytical pressure profiles of Equation \ref{pre} are plotted as a function of radius and time. Figure is produced after~\cite{Nick2023} with the same values on the axes, but (the present figure is) in dimensionless forms, for comparison purpose.}\label{figure_ana_pre}
\end{figure}

The fluid mechanics of impacting liquids is analytically described by Hao \textit{et al.}~\cite{Hao_ana}, where the authors assumed the flow as an incompressible and inviscid axisymmetric ideal fluid field, and the impact surface as a rigid material. Hao \textit{et al.} derived an analytical solution for the droplet impact pressure on the solid surface as:
\begin{align}
  p(r,t,\theta) &= - \frac{2}{\pi^2} {\left( cos(\theta) - \frac{r}{\sqrt{a(t)^2-r^2}} \; sin(\theta) \right)}^2 \nonumber\\
&+ \frac{2}{\pi} \frac{a(t)a'(t)sin(\theta)}{\sqrt{a(t)^2-r^2}} \;\;\;\;\;\;\;\; \text{for $r \leq r_{max}(t,\theta)$} \label{sol_pre}
\end{align}
and zero elsewhere, where $r=\sqrt{x^2+y^2}$ is the radial distance from the impact surface center in cylindrical coordinate (see Figure \ref{ring}a but in Cartesian coordinate), $t$ is time starting from the moment of impacts, $\theta$ is the impact angle from vertical axis, $a$ is the radial position of the spreading wet radius, $r_{max}$ is the radial position where pressure is maximum, and all lower-case letters denote dimensionless forms of corresponding upper-case variables (except density) with respect to the droplet initial radius $R_0$, the impact velocity $U_0$ and the liquid density $\rho_l$. In the context of liquid droplet impact at normal angle, we have $a(t)=\sqrt{3t}$~\cite{Wagner} and  $\theta=\pi/2$, which give the surface pressure:
\begin{equation}
  p(r,t) = - \frac{2}{\pi^2} \frac{r^2}{3t-r^2} \; + \frac{3}{\pi} \frac{1}{\sqrt{3t-r^2}} \;\;\;\;\;\;\;\; \text{for $r \leq r_{max}(t)$} 
\label{pre}
\end{equation}
and
\begin{equation}
  r_{max}(t)=\sqrt{3t-\frac{16t}{3 \pi^2}}
\label{r_max}
\end{equation}
The dimensionless values of $p(r,t)$ as a function of radius and time are plotted in Figure \ref{figure_ana_pre}. Readers are referred to Figure 2 of Nick \textit{et al.}~\cite{Nick2023} for comparison, where a comparable figure calculated from a computational fluid dynamics (CFD) model exists with same values on the axes\footnote{except their droplet impact happens at $T = 1$ \unit{\micro\second}} and (their) pressure color bars in dimensional forms. In the current study, Equations \ref{pre} and \ref{r_max} are the analytical pressure solutions on the contact surface of the solid material upon liquid droplet impact.

\subsection{Numerical material simulation}
\label{2.2}
\begin{figure*}
\centering
\includegraphics[width=16 cm]{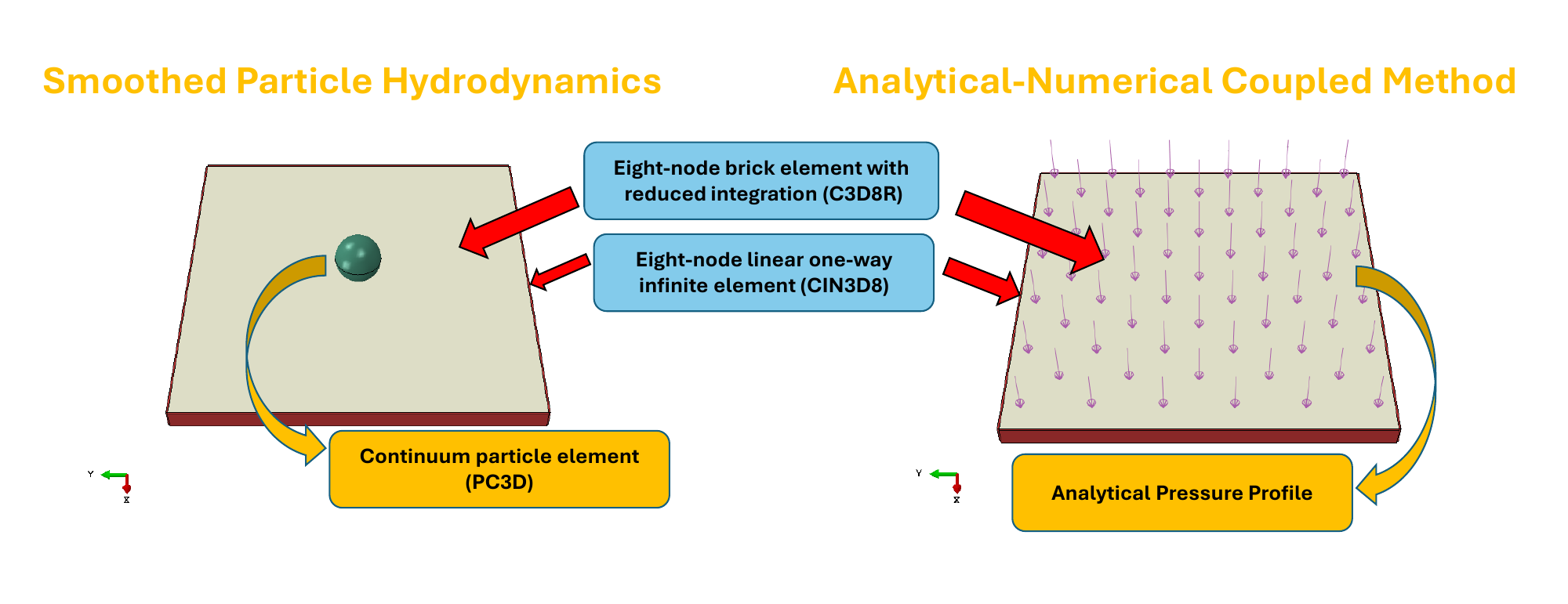}	
\caption{Required FE elements (with element codes in ABAQUS) and loading conditions in SPH (left) and ANCM (right) models~\cite{Hao_ANCM}. The figure is for illustration purpose and the solid material dimensions are not to scale.}\label{figure_ANCM}
\end{figure*}

Soft solid material is modeled in commercial software ABAQUS\raisebox{1ex}{\scriptsize \textregistered} using three-dimensional Finite Element (FE) analysis. Solid materials are modeled by the eight-node brick element with reduced integration (C3D8R) with a layer of one-way infinite elements (CIN3D8) on four sides and bottom to get rid of wave reflections due to the finite geometry. The development of one-way fluid-structure interaction coupling framework~\cite{Hao_ANCM} combines the analytical pressure solution of Equations \ref{pre} and \ref{r_max} as a time- and spatial-dependent loading on the surface of the solid material to model the impact loading of a liquid droplet. This user-defined loading is achieved by the VDLOAD-subroutine in ABAQUS for explicit dynamic analysis. Figure \ref{figure_ANCM} displays the geometry and FE elements of the study. Four sides and bottom of the geometry are fixed by encaster boundary conditions to model the geometric constraints of the simulated material sections in (relatively) semi-infinite material domains in real engineering applications. For comparison, the traditional Smoothed Particle Hydrodynamics (SPH) method~\cite{Hao_ANCM} is shown on the left, in which, liquid droplet is modeled by continuum particle elements (PC3D) in FE analysis.

\begin{figure*}[t]
\centering
\includegraphics[width=16 cm]{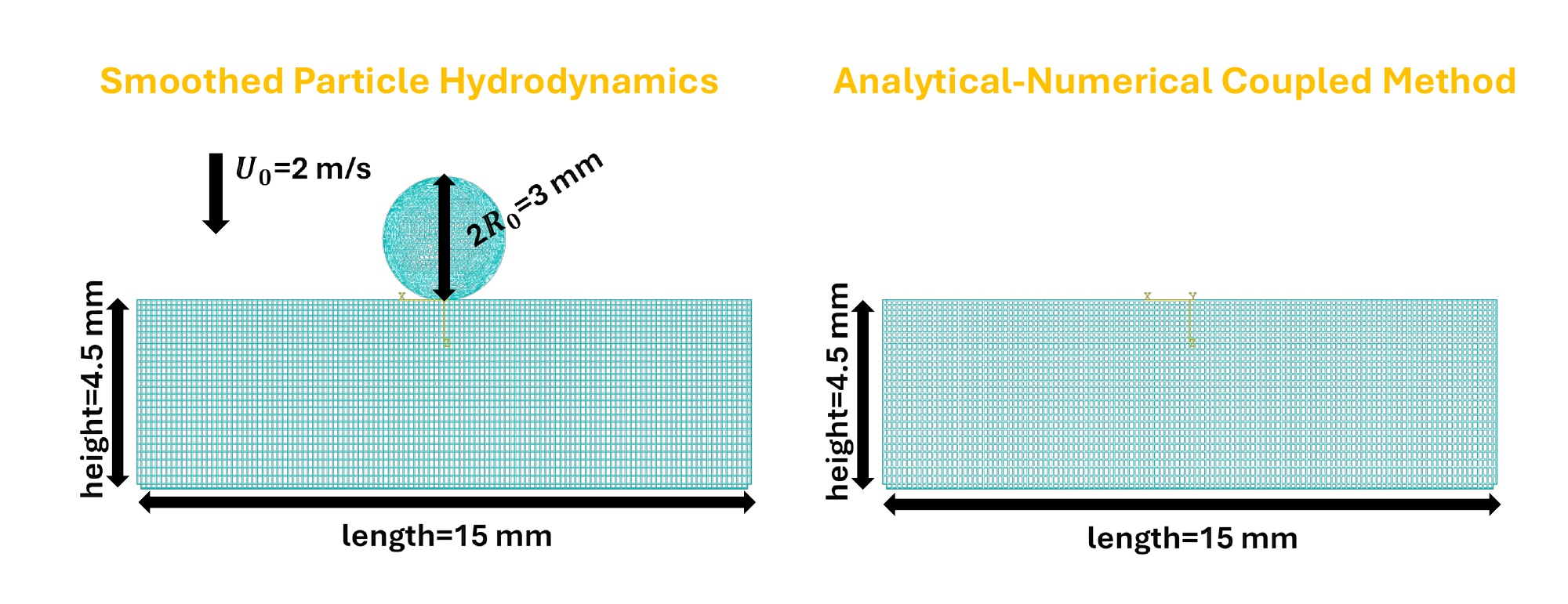}	
\caption{Numerical FE mesh and simulation parameters in SPH (left) and ANCM (right) models. The same mesh resolution has been used for the solid materials in SPH (left) and ANCM (right) models.}\label{figure_mesh}
\end{figure*}

The three dimensional FE model is solved on a substrate of size $15\text{ mm (length)} \times 15\text{ mm (width)}\times 4.5\text{ mm (height)}$ with total simulation time of $3$ \unit{\milli\second} at a timestep of $10$ \unit{\micro\second}. The timestep is selected based on the guidance that peak pressure force shall be captured in no less than ten time steps~\cite{Leon}, and our sensitivity check indeed shows no results difference at finer timesteps. The substrate is bias-meshed with finer mesh towards the impact surface (mesh size $125$ \unit{\micro\meter}) and coarser at bottom (mesh size $200$ \unit{\micro\meter}). The total number of elements (including the one-way infinite layers) is 0.43 million, which is in line with the order of FE element number for droplet impact simulation in the literature~\cite{XJ}. In the SPH model, the extra droplet geometry is entirely meshed with the fine grids (mesh size $125$ \unit{\micro\meter}) as each node contacts the surface. This adds up the total number of elements (including the one-way infinite layers) to 0.46 million. The settings of the numerical model for the present study result in the mesh geometries of Figure \ref{figure_mesh}. The element sizes of the numerical model are selected based on reasonable computing power and model accuracy, in the way that elements in contact have similar sizes. In the previous study~\cite{Hao_ANCM}, grid-independence test of the developed numerical model was performed on a uniform mesh (for both droplet and solid material) of even finer resolution (mesh size $89$ \unit{\micro\meter}, equivalent to 3\% of droplet diameter). Applying such mesh resolution to the solid geometry of the current study requires total number of elements over 1.6 million, however the obtained SPH simulation results showed negligible difference to results from the mesh resolution defined here. Therefore, the current mesh configuration is retained. Typical simulation of the defined ANCM model takes about 12 min to solve on a AMD Ryzen Threadripper PRO 5975WX\raisebox{1ex}{\scriptsize TM} processor and 270 GB of RAM running in parallel at 16 cores. The fully numerical method of SPH takes about 1540 min at the same computing resources. This highlights the fact that the extra 0.03 million droplet mesh (and the computation of the contact algorithm) takes the majority of the computing power as the liquid droplet undergoes the largest deformation during the impact.

\begin{table}
\caption{Summary of material properties of the urethane gel block (top) and impact parameters of the liquid silicone oil droplet (bottom) used in the experiments of \cite{Yuto2024}.}\label{table_mat}
\centering
\begin{tabular}{ccc} 
\toprule
\textbf{Young's Mod. $E$} &  \textbf{Den. $\rho_s$}&  \textbf{Poisson's Rat. $\nu$}\\
\midrule
47,400 (\unit{\pascal})	& 1,000 (\unit{\kilogram\per\cubic\meter})& 0.499\\
\bottomrule
\toprule
\textbf{Drop. Rad. $R_0$} &  \textbf{Imp. Vel. $U_0$} & \textbf{Den. $\rho_l$}\\
\midrule
1.5 (\unit{\milli\meter})	& 2  (\unit{\meter\per\second})& 815.546 (\unit{\kilogram\per\cubic\meter})\\
\bottomrule
\end{tabular}
\end{table}

\begin{table*}[t]
\caption{Summary of the five cases of parametric studies on material stiffness, while other material properties and impact parameters are kept the same as in Table \ref{table_mat}. Case $1$ is the reference case of the experiment in \cite{Yuto2024}.}\label{table_para}
\centering
\begin{tabular}{cccccc} 
\toprule
 & \textbf{Case 1} & \textbf{Case 2} & \textbf{Case 3} & \textbf{Case 4} &  \textbf{Case 5}\\
\midrule
\textbf{Young's Modulus $E$} & 47,400 (\unit{\pascal})&20,000 (\unit{\pascal})&12,000 (\unit{\pascal})&10,000 (\unit{\pascal})&4,740 (\unit{\pascal})\\
\bottomrule
\label{table_para}
\end{tabular}
\end{table*}

The developed analytical-numerical coupled model (ANCM) has been validated~\cite{Hao_ANCM} by the experimental data using aluminum plate~\cite{XJ}. Since aluminum material is stiff, it is reasonable to satisfy the assumption of rigid surface in the development of the analytical pressure solution. In the current study, however, we apply the ANCM model on soft materials (Young's modulus less than $1$ \unit{\mega\pascal}), which apparently do not satisfy the theoretical assumptions of rigid surface. In this way, we intend to explore the applicability limit of the developed ANCM model on soft materials. A urethane gel block (urethane gel phantom, Exseal Co., Ltd. Gifu, Japan) is used in the current study as the soft solid material impacted by silicone oil droplets of kinematic viscosity $1$ cSt. The silicone oil droplet is selected to reasonably satisfy the inviscid assumption of the analytical pressure solution. Solid material properties (a linear elastic model), liquid droplet geometry and properties, and impact parameters are listed in Table \ref{table_mat}. Experimental data is averaged from three tests at the same impact velocity. Readers are referred to Yokoyama \textit{et al.}~\cite{Yuto2024} for further experimental details. Parametric studies on the material elasticity are carried out by artificially varying the solid material Young's modulus in numerical simulations while keeping other parameters unchanged. Table \ref{table_para} summarizes the five cases studied, where case $1$ is the reference case as in experiments. Each case will be numerically simulated using both the conventional SPH model and the developed ANCM for comparison, and case $1$ will also be compared to experiments.

\section{Results and Discussion}
\label{results}
\subsection{Model evaluation with experimental data}
\label{3.1}
\begin{figure}
\centering
\includegraphics[width=9.5 cm]{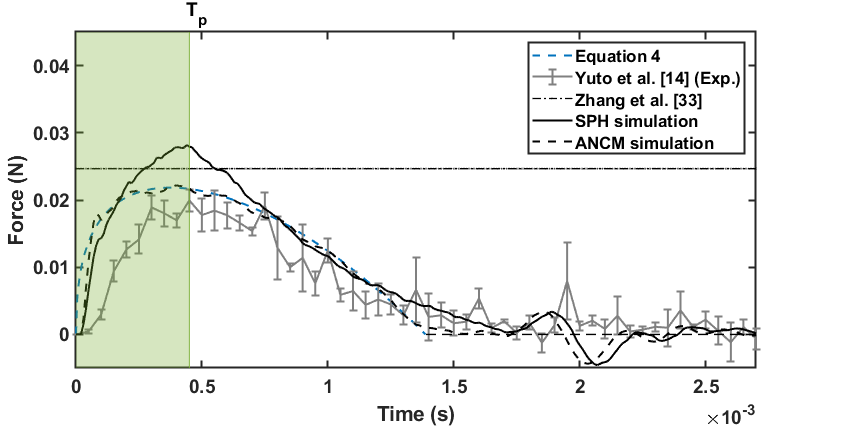}	
\caption{The temporal evolutions of the impact force on the solid surface, from the experiment in~\cite{Yuto2024} and the numerical simulations of SPH and ANCM models, are compared. The error bar indicates one standard deviation of three tests in experiments. The analytical solution of Equation \ref{temp1}, which is the impact force under the liquid droplet assuming a rigid surface, is also superimposed. The experimental peak force happens at $T_p=0.45\times10^{-3}$ \unit{\second}. The empirical formula in~\cite{Zhang} estimates the peak force as $F_{emp,p}=0.0247$ \unit{\newton}.}\label{force_ref}
\end{figure}

Simulation results of the two numerical models for silicone oil droplets of viscosity $1$ cSt impacting upon phantom substrate of Young's modulus $47,400$ \unit{\pascal} are validated against experimental data. The total contact force on the surface of the substrate as a function of time, which is in the direction of impact, is compared to measurements in Figure \ref{force_ref}. As can be seen, the computed contact forces from both the SPH and ANCM numerical models agree well with the measurements. They all predict a curved contact force with peak at time of around $0.5$ \unit{\milli\second}, and the contact force then gradually decays to zero as the liquid droplet spreads on the surface. It shall be noticed that the contact force from ANCM is output at the material surface from FE analysis, which is not necessarily the same as the input analytical impact loading on the substrate surface. To better understand the difference, we analytically integrate the pressure solution of Equation \ref{pre} on the rigid solid surface, and obtain the contact force:
\begin{align}
    f(t) &=\int_0^\infty \int_0^{2\pi} p(r,t) r\;d\alpha \;dr\nonumber\\
    &=\int_0^{r_{max}(t)} \int_0^{2\pi} p(r,t) r\;d\alpha \;dr\nonumber\\
    &=2\pi\int_0^{r_{max}(t)} p(r,t) r\;dr
    \label{temp1}
\end{align}
where $(r,\alpha)$ is the polar coordinate on the rigid surface with the origin at the impact center, and $ p(r,t)$ is from Equation \ref{pre} and $r_{max}(t)$ is from Equation \ref{r_max}. Explicit solutions of Equation \ref{temp1} are referred to Hao \textit{et al.}~\cite{Hao_ANCM}. The obtained analytical force by Equation \ref{temp1} is compared with the experimental data in Figure \ref{force_ref}. An empirical peak impact force $F_{emp,p}=3.36\rho U_0^2R_0^2$~\cite{Zhang} for liquid droplet impacts is also superimposed. As can be seen, the empirical peak impact force slightly overpredicts the peak force of the measurements, while the analytical solution agrees well with the experiments, \emph{i.e.} mostly within the uncertainty range except a few frames at the beginning of the impact. This agreement is well-reflected in the material part of the coupling framework that, in Figure \ref{force_ref}, the FE numerical simulation from ANCM reproduces the particular shape of the analytical impact force (this is not necessarily always hold, as we shall see), and provides contact force that matches with the experimental impact force in the exact same way. While it is found that the SPH model overpredicts the peak impact force measurements. It is worth-noting that the analytical impact force vanishes at time $1.39$ \unit{\milli\second} in an unsmooth way, while FE simulation of ANCM smoothens the analytical impact loading and produces the smooth impact force curve in Figure \ref{force_ref}.

\begin{figure*}
    \centering
    \begin{subfigure}[b]{0.48\textwidth}
        \centering
        \includegraphics[width=7 cm]{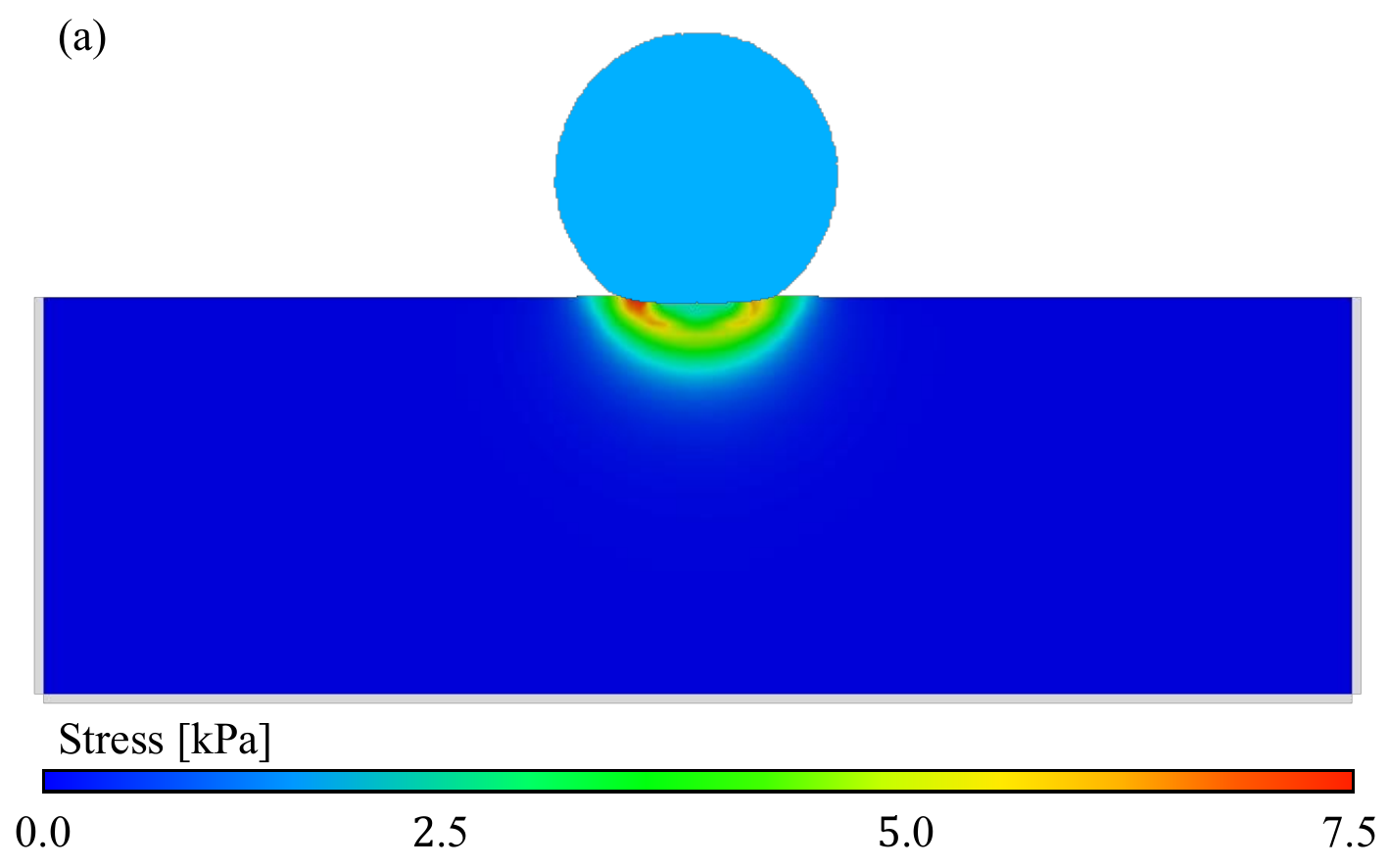}
    \end{subfigure}
    \begin{subfigure}[b]{0.48\textwidth}
        \centering
        \includegraphics[width=7 cm]{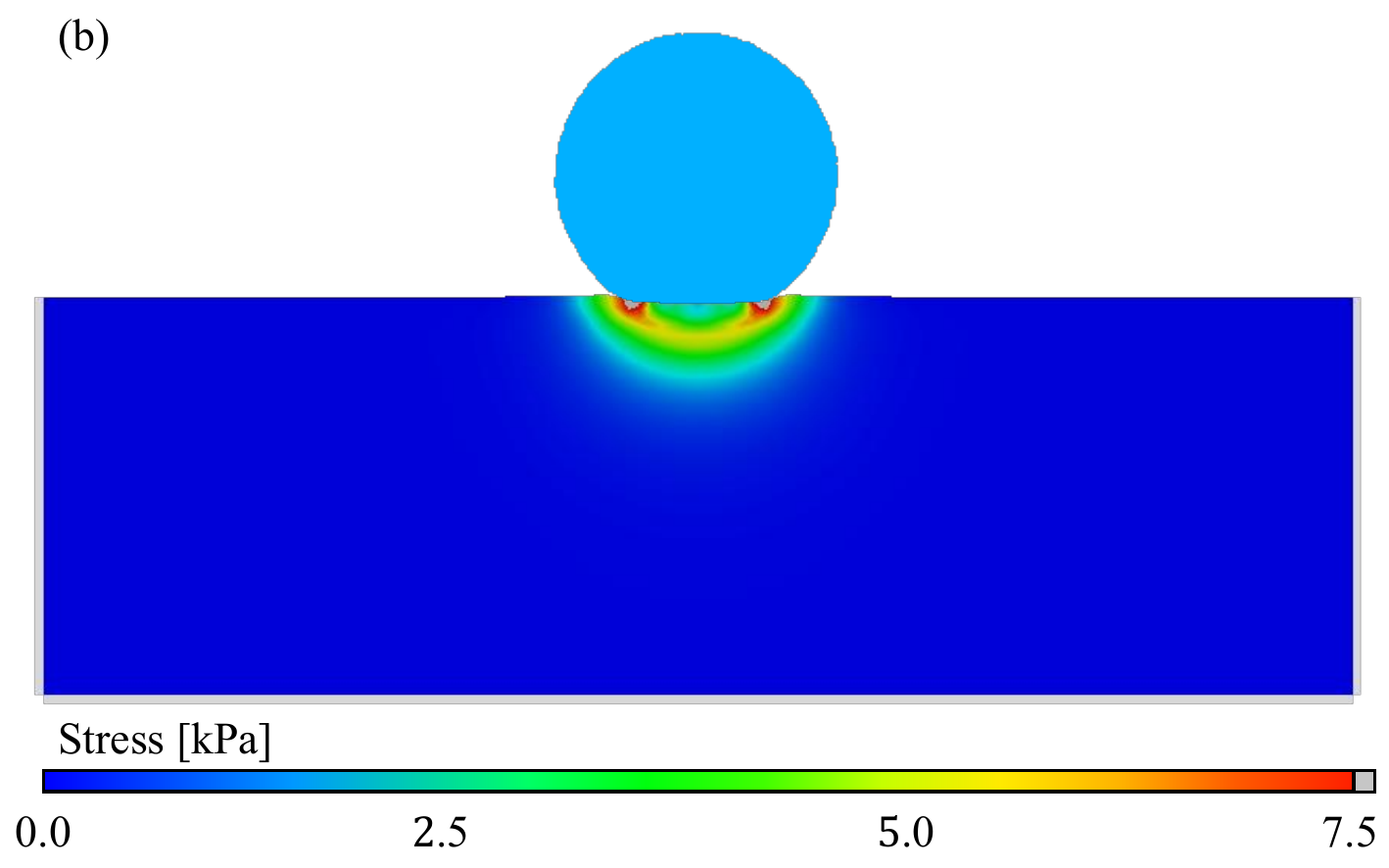}
    \end{subfigure}
    \begin{subfigure}[b]{0.48\textwidth}
        \centering
        \includegraphics[width=7 cm]{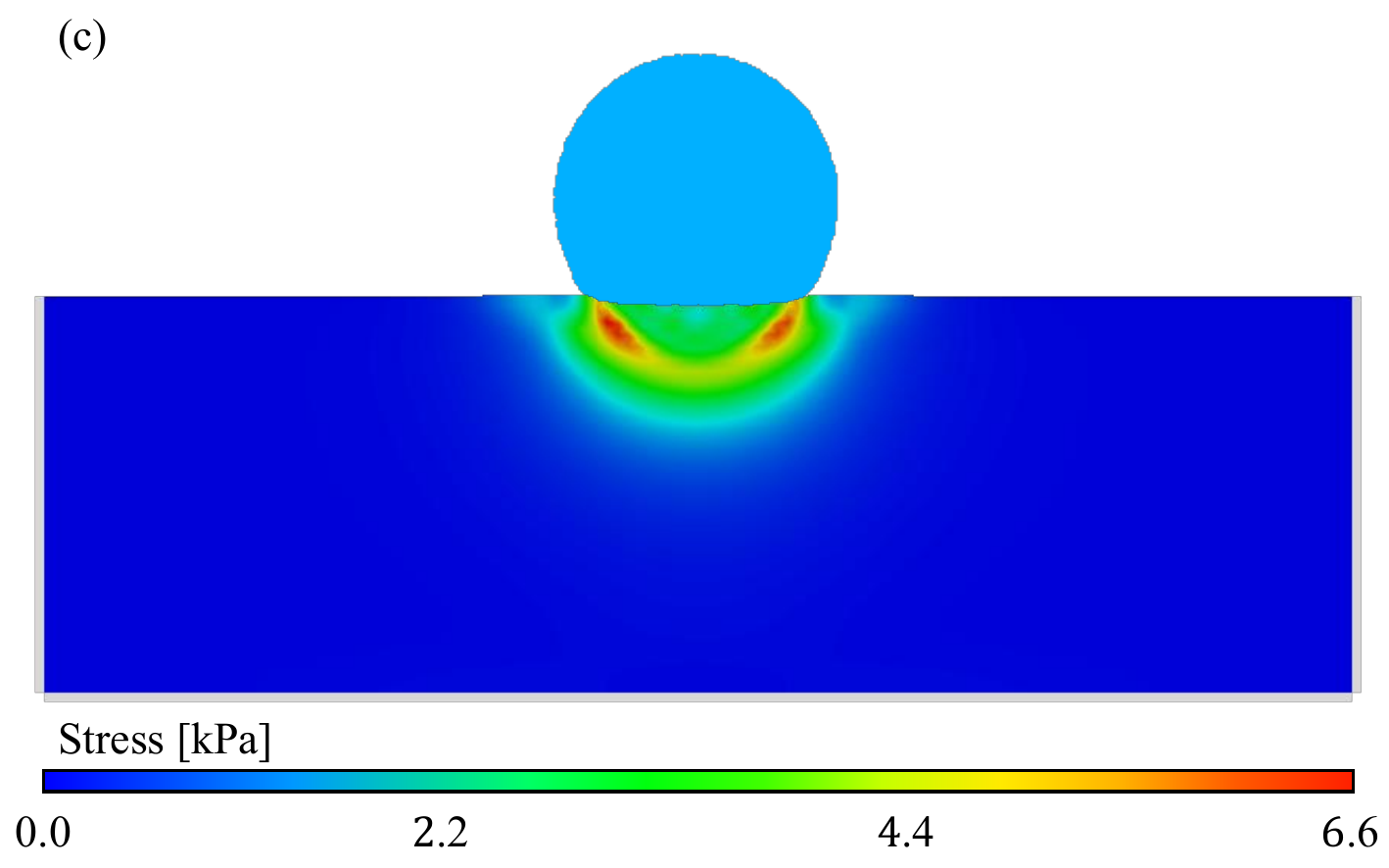}
    \end{subfigure}
    \begin{subfigure}[b]{0.48\textwidth}
        \centering
        \includegraphics[width=7 cm]{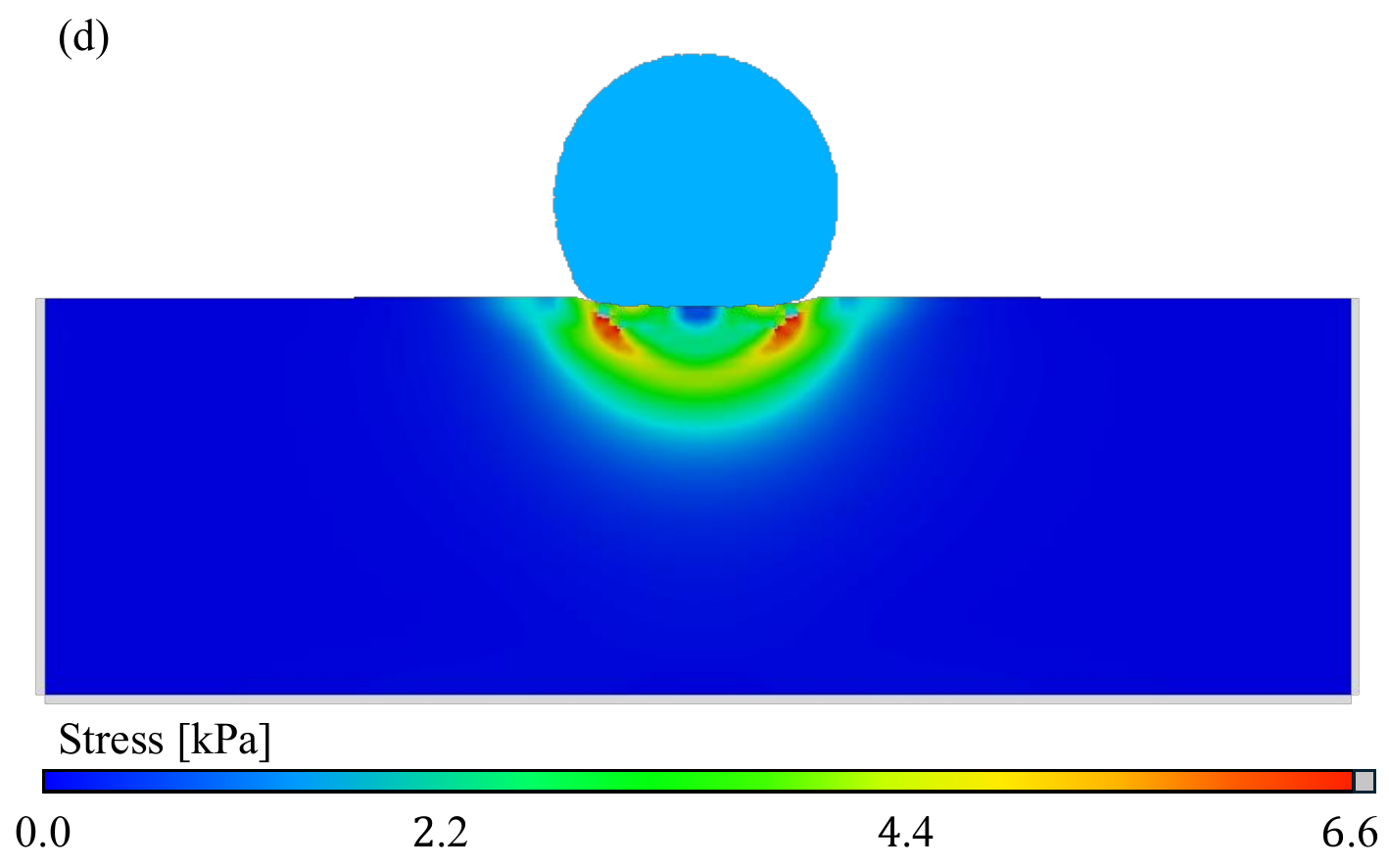}
    \end{subfigure}
    \begin{subfigure}[b]{0.48\textwidth}
        \centering
        \includegraphics[width=7 cm]{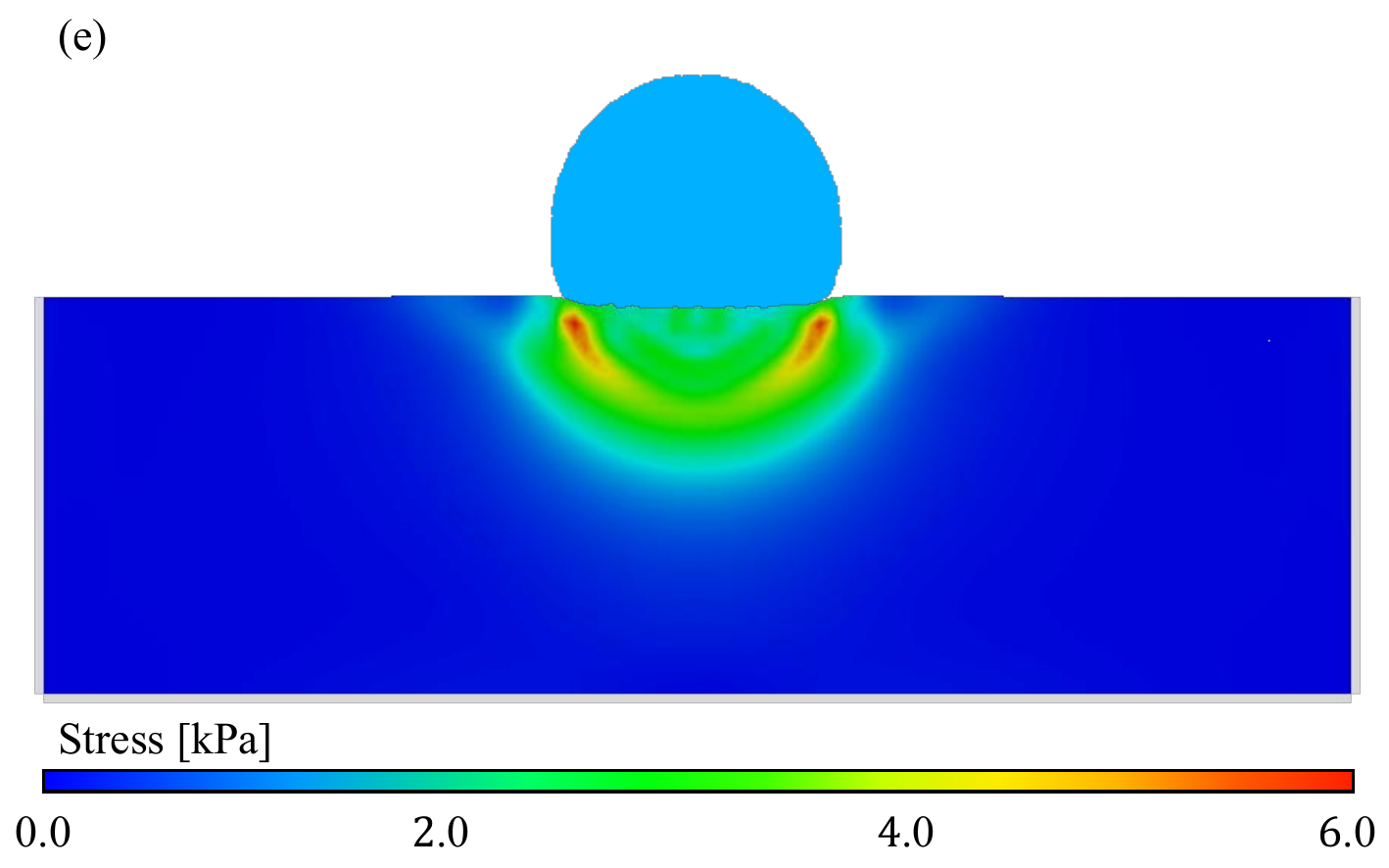}
    \end{subfigure}
    \begin{subfigure}[b]{0.48\textwidth}
        \centering
        \includegraphics[width=7 cm]{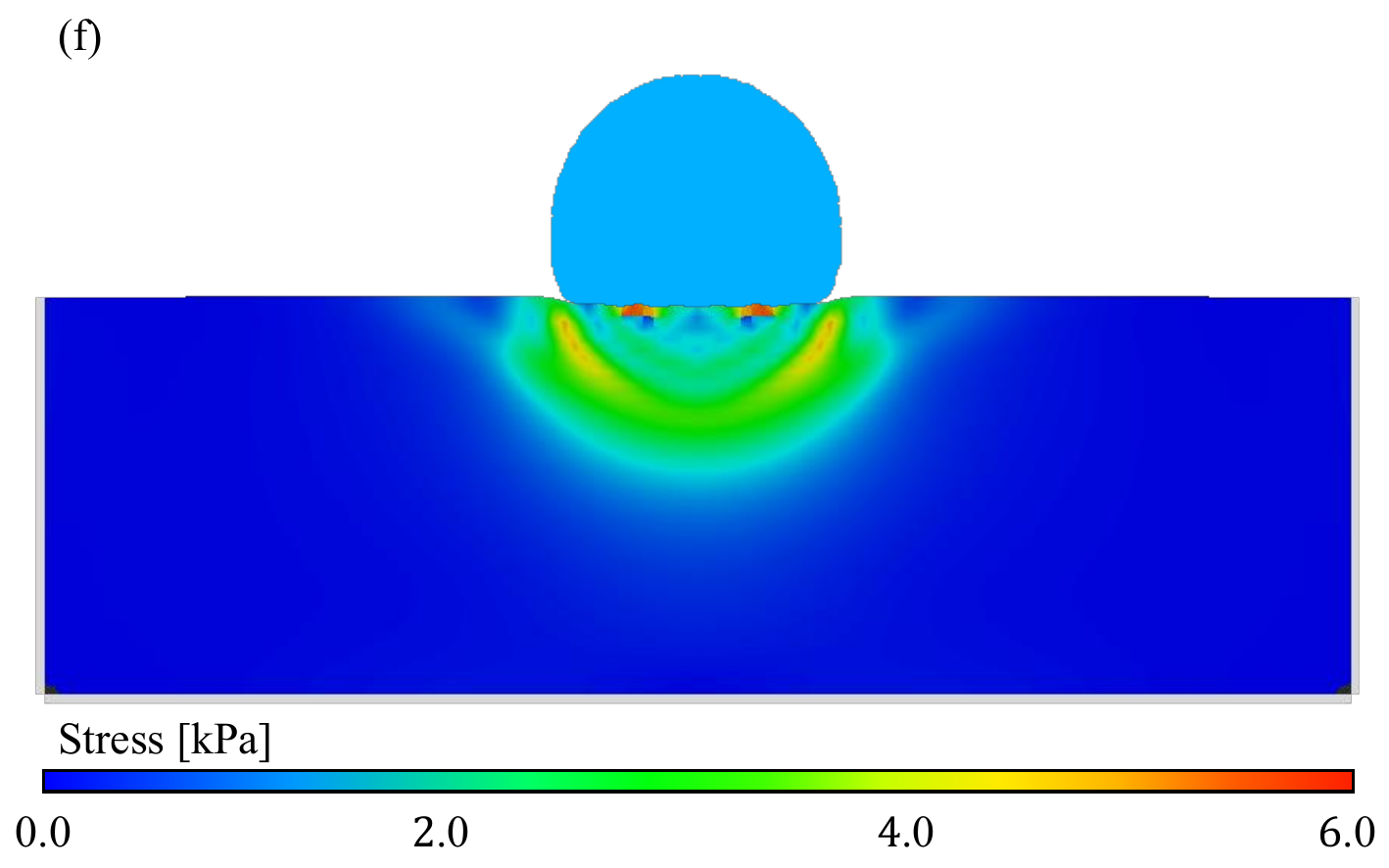}
    \end{subfigure}
    \begin{subfigure}[b]{0.48\textwidth}
        \centering
        \includegraphics[width=7 cm]{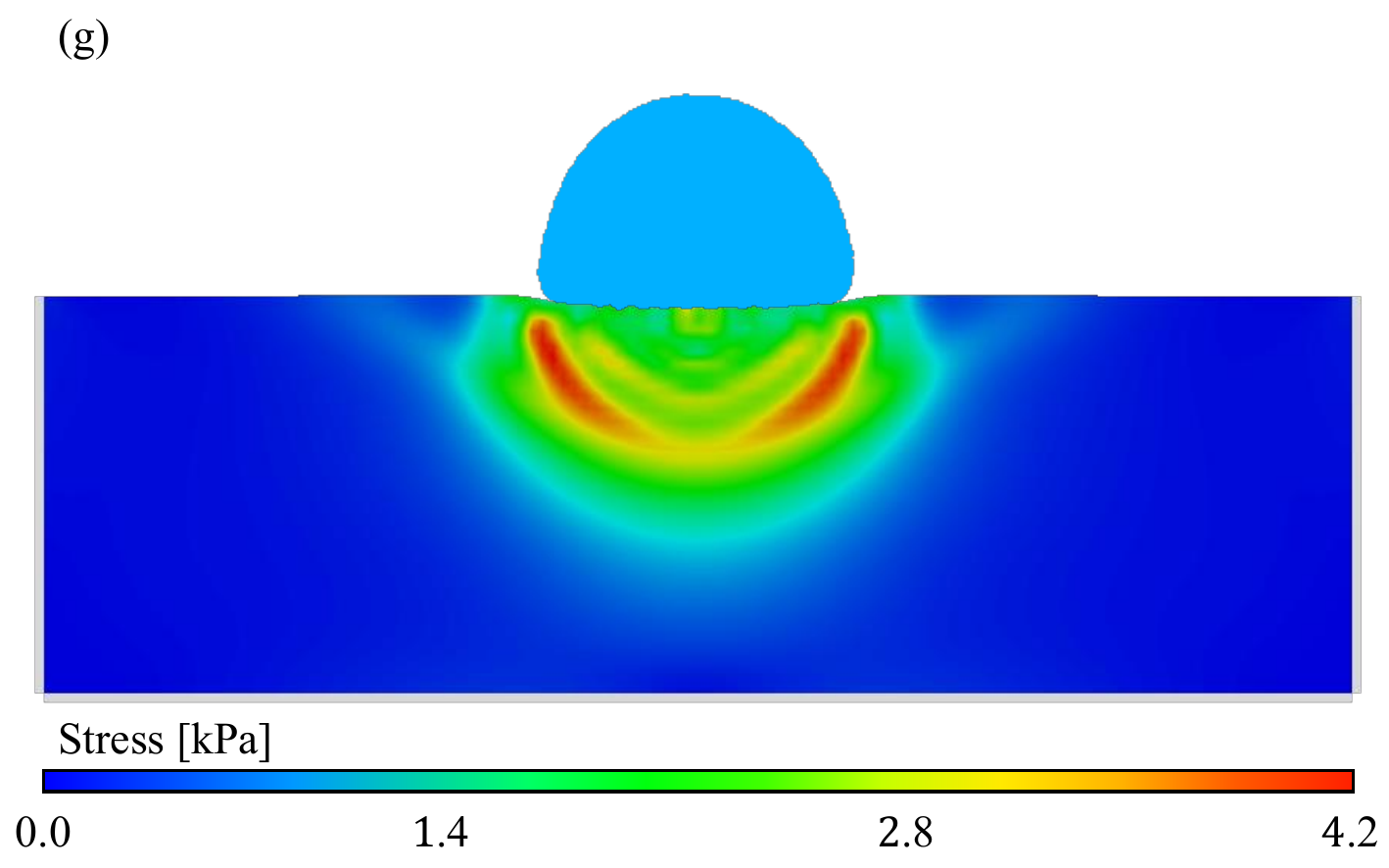}
    \end{subfigure}
    \begin{subfigure}[b]{0.48\textwidth}
        \centering
        \includegraphics[width=7 cm]{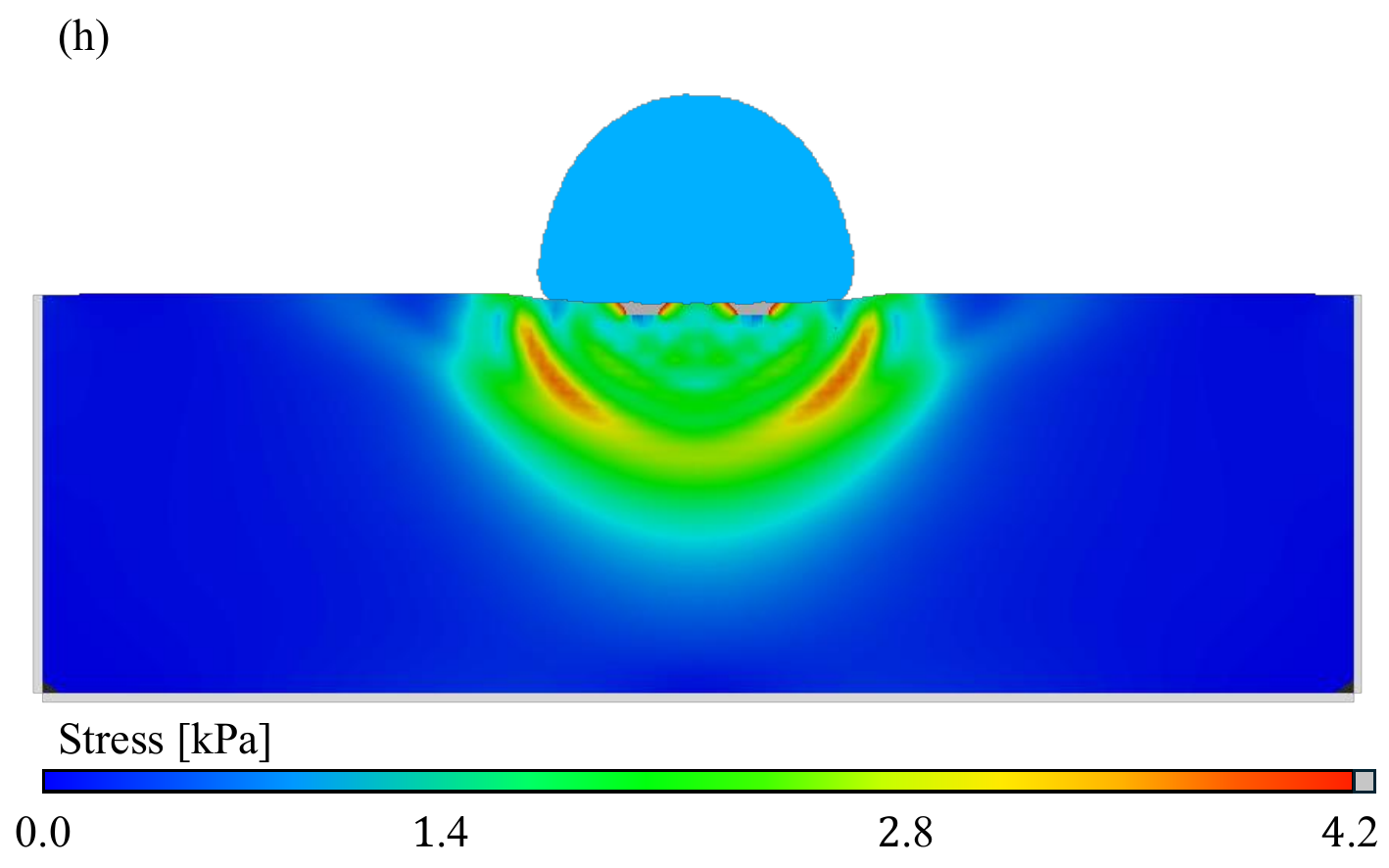}
    \end{subfigure}
    \begin{subfigure}[b]{0.48\textwidth}
        \centering
        \includegraphics[width=7 cm]{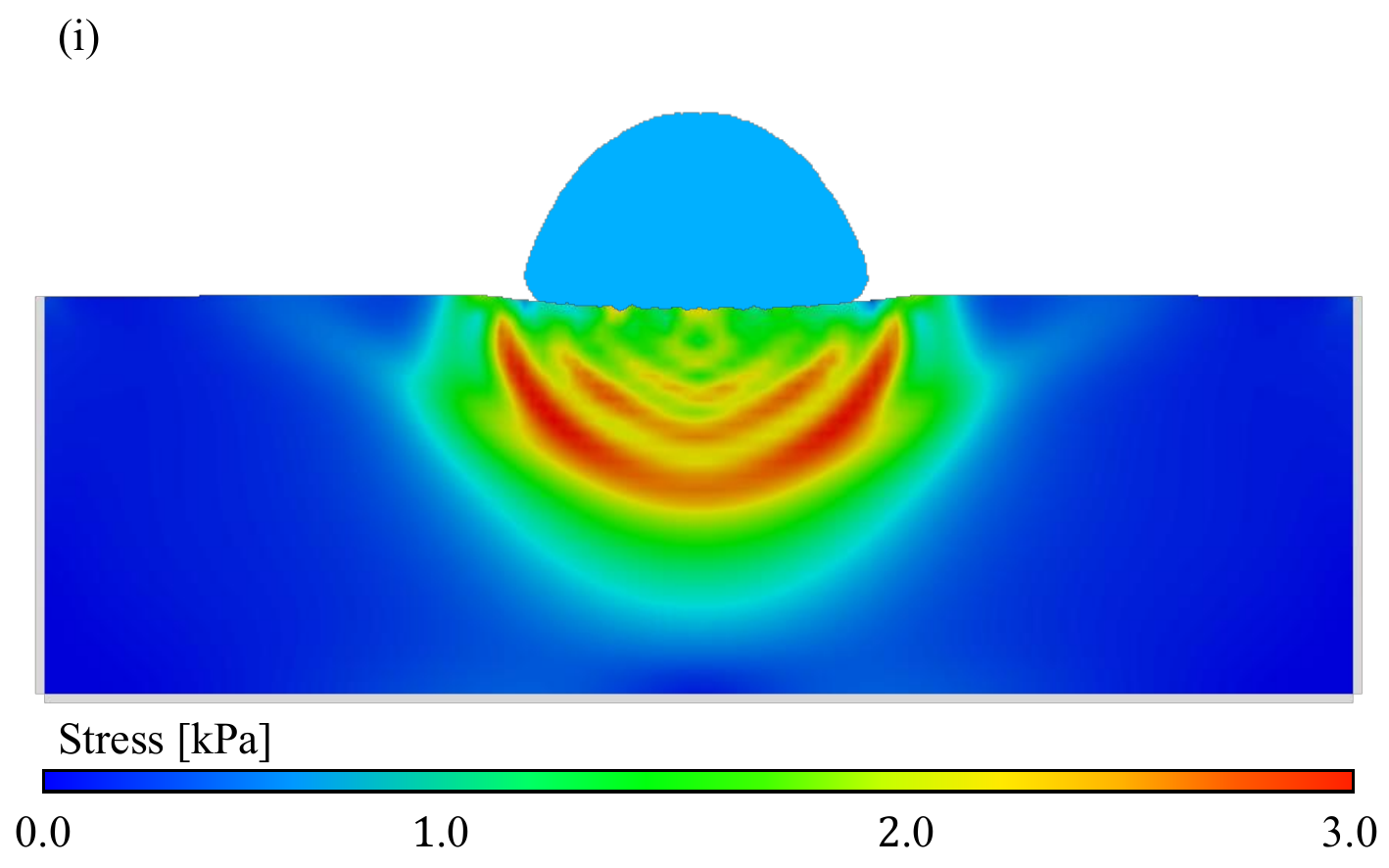}
    \end{subfigure}
    \begin{subfigure}[b]{0.48\textwidth}
        \centering
        \includegraphics[width=7 cm]{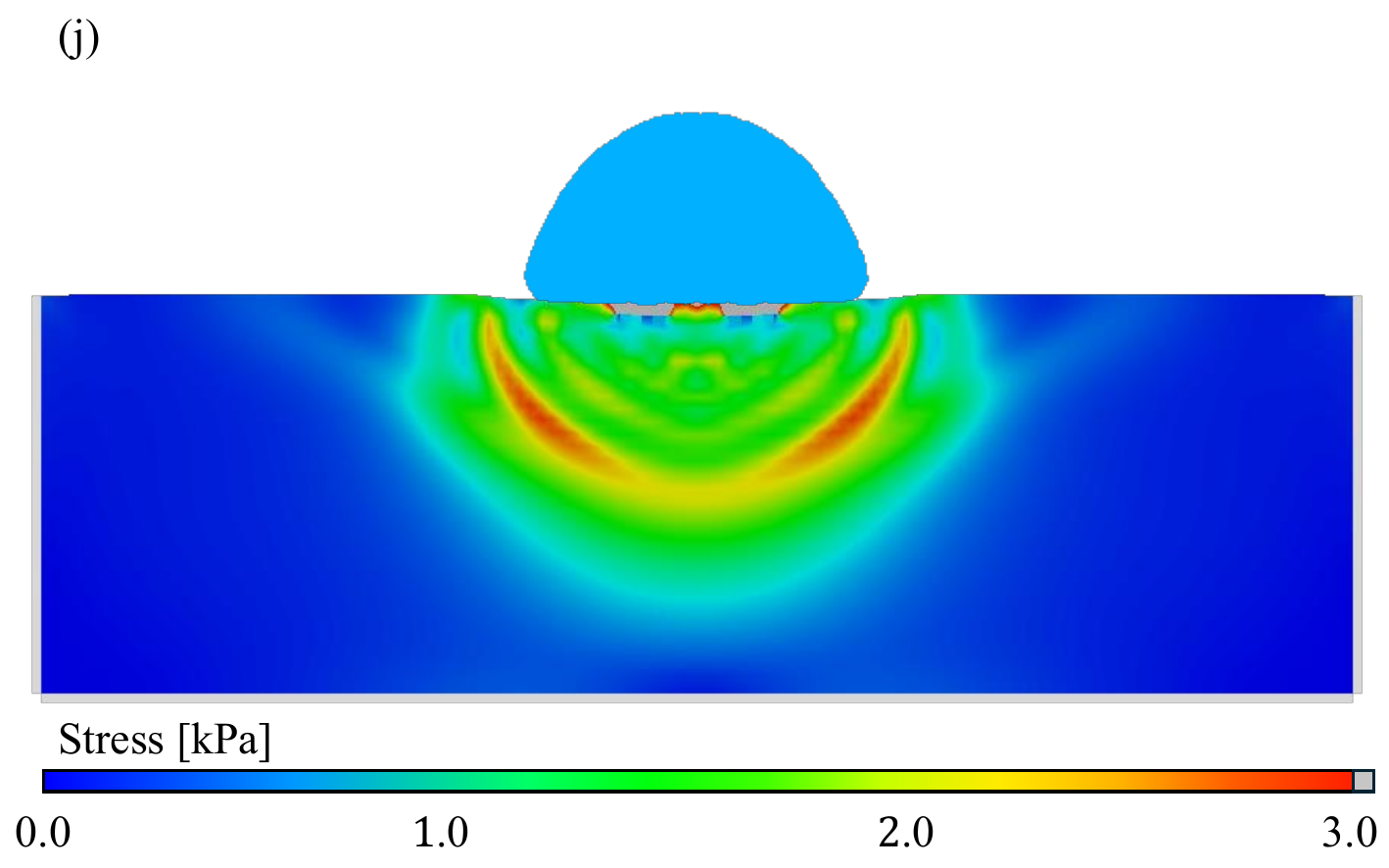}
    \end{subfigure}
    \caption{Spatial distributions of von Mises stress field inside the surface material obtained from the SPH (left column) and ANCM (right column) simulations are shown at impact times of $120$ \unit{\micro\second} (a, b), $240$ \unit{\micro\second} (c, d), $360$ \unit{\micro\second} (e, f), $480$ \unit{\micro\second} (g, h) and $590$ \unit{\micro\second} (i, j). Each subfigure shows the front view with a view cut at the $y$-$z$ plane of symmetry, \emph{i.e.} the same viewing position as in Figure \ref{figure_mesh}. The same colorbar limits have been used for the SPH (left column) and ANCM (right column) results at each impact time. For illustration purpose, droplet configurations are added to the ANCM results from the SPH simulation at corresponding times. Note that the droplet color is independent of the stress colormap.}
\label{result_ref}
\end{figure*}

The dynamic von Mises stress field obtained from SPH (left column) and ANCM (right column) is shown in Figure \ref{result_ref}. The stress fields (and corresponding droplet configurations) are plotted at impact times of $120$ \unit{\micro\second}, $240$ \unit{\micro\second}, $360$ \unit{\micro\second}, $480$ \unit{\micro\second}, $590$ \unit{\micro\second} respectively. It can be seen that the obtained stress fields from the two methods agree well in trends and stress magnitudes. In particular, the spatial and temporal expanding of the stressed area, and the behaviors of the local high-stress ring strips\textemdash generated around the contact radius of the droplet\textemdash agree well for the two methods. The subtle difference between the two models is the stress behaviour at the impact surface that ANCM predicts relatively larger stress at the periphery but lesser stress at the impact center (see Figure \ref{result_ref}a and Figure \ref{result_ref}b). Propagating into the substrate with time, this facial stress difference leads to the uniform ring strips in SPH (Figure\ref{result_ref}i) but the non-uniform ring strips in ANCM (Figure \ref{result_ref}j).

\begin{figure*}
\centering
\includegraphics[width=16 cm]{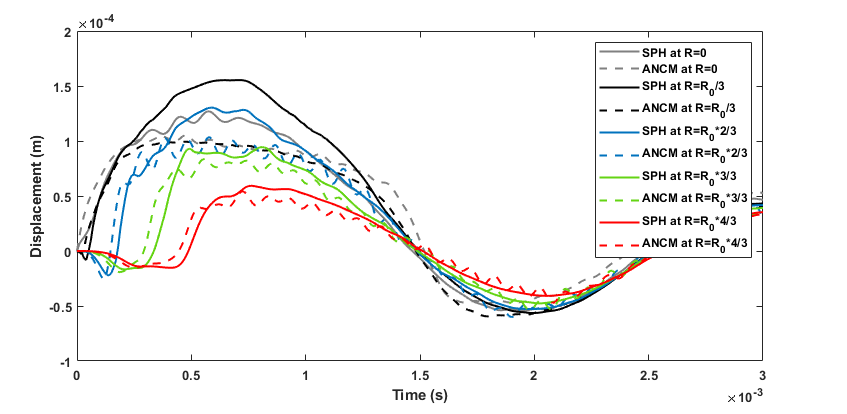}	
\caption{The (vertical) displacements are shown at five locations on the surface of the solid material as a function of time after the start of the droplet impact in SPH (solid lines) and ANCM (dashed lines) models, of radii $0$ \unit{\milli\meter}, $0.5$ \unit{\milli\meter} ($R=R_0/3$), $1.0$ \unit{\milli\meter} ($R=R_0*2/3$), $1.5$ \unit{\milli\meter} ($R=R_0*3/3$) and $2.0$ \unit{\milli\meter} ($R=R_0*4/3$) respectively. Locations and color codes correspond to the points of Figure \ref{def_loc}. $R_0=1.5$ \unit{\milli\meter} is the initial droplet radius.}\label{def_ref}
\end{figure*}

\begin{figure}
\centering
\includegraphics[width=9.5 cm]{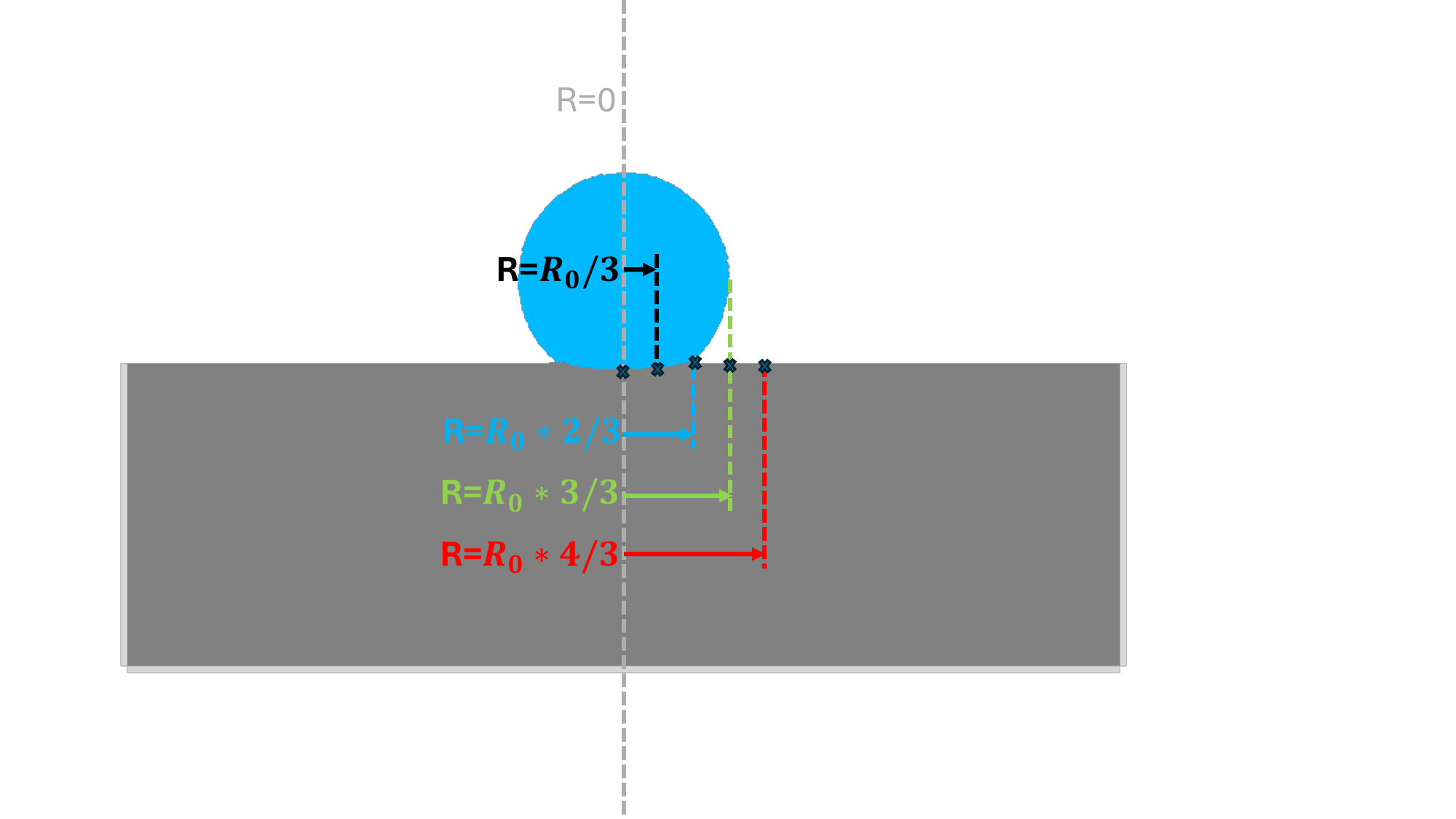}	
\caption{Figure illustrates the probed five locations on the surface of the solid material, of radii $0$ \unit{\milli\meter} (R=$0$), $0.5$ \unit{\milli\meter} (R=$R_0/3$), $1.0$ \unit{\milli\meter} (R=$R_0*2/3$), $1.5$ \unit{\milli\meter} (R=$R_0*3/3$) and $2.0$ \unit{\milli\meter} (R=$R_0*4/3$), respectively. The gray dashed line represents the central axis. For illustration purpose, the figure is generated from the SPH simulation at impact time $1.2$ \unit{\micro\second}. $R_0=1.5$ \unit{\milli\meter} is the initial droplet radius.}\label{def_loc}
\end{figure}

Also in Figure \ref{result_ref}, particular interest is given to the surface deformation of the substrate in the two numerical methods. To facilitate comparison, we add droplet configurations in ANCM results from the SPH simulation at corresponding times. As can be seen, from small dimples at the very beginning of the impact (Figure \ref{result_ref}a, \ref{result_ref}b) to large wavy surface (Figure \ref{result_ref}i, \ref{result_ref}j), SPH and ANCM predict with satisfactory consistency the surface deformation. Probing into the details, Figure \ref{def_ref} presents the temporal evolution of the vertical displacements at five locations on the surface of the solid material, of radii $0$ \unit{\milli\meter} (i.e. impact center), $0.5$ \unit{\milli\meter}, $1.0$ \unit{\milli\meter}, $1.5$ \unit{\milli\meter} and $2.0$ \unit{\milli\meter} respectively (corresponding to locations shown in Figure \ref{def_loc}), from the simulation results of SPH and ANCM. It is noted that the $z$-axis is selected as the direction of impact, so positive vertical displacements indicate depression of the surface. For general trends, the substrate surface is depressed in the middle, followed by reflection upwards, and then repeats the wavy motion elastically. While away from the impact center, surface is first elevated due to the `squeeze effect' of the depression at the impact center\textemdash so-called retrograde motion~\cite{Miller, Kim}, and is then followed by the same elastic wavy motions. Specifically, at radial position of two-thirds of initial droplet radius ($1.0$ \unit{\milli\meter}), there is the largest surface deformation with magnitude of around $0.017R_0$ ($25$ \unit{\micro\meter}). While at radial position of one-third of initial droplet radius ($0.5$ \unit{\milli\meter}), there is the largest surface deformation as the surface is depressed, with magnitude of around $0.1R_0$ ($155$ \unit{\micro\meter}). Thereby, the substrate undergoes the maximum surface deformation deflections between radial positions of one-third ($0.5$ \unit{\milli\meter}) and two-thirds ($1.0$ \unit{\milli\meter}) of initial droplet radius as a ring-shape pattern. This agrees with the experimental observations of ring-shape damage patterns in the literature~\cite{Field2}. Comparing the results from SPH and ANCM, the trends at early impact time are in-line with the impact force difference (Figure \ref{force_ref}) that ANCM predicts a slightly earlier peak while the SPH overpredicts the peak with higher magnitude. This leads to the peak displacement difference between the results of SPH and ANCM, although the displacement curves from the two methods are consistent. Nevertheless, the difference is reduced as the distance from the center increases, and the calculated deformations at $R\geq R_0$ from the two methods are matched very well.

\subsection{Parametric study}
\label{3.2}
After the evaluation of the models for the reference case (Case $1$) with the experimental data, we regard the SPH model to be valid as a benchmark for the following analysis. Therefore, we compare the SPH results with the ANCM predictions (which have not been validated on softer materials) for softer materials associated with lower Young's modulus (see Table \ref{table_para}). It is reminded that the purpose of this parametric study is to find the limitation and possible failure of the ANCM model in engineering simulations as materials deviate further from the rigid surface assumption of the analytical solution.

\subsubsection{Total contact force}\label{TCF}
\begin{figure}[t]
\centering
\includegraphics[width=9.5 cm]{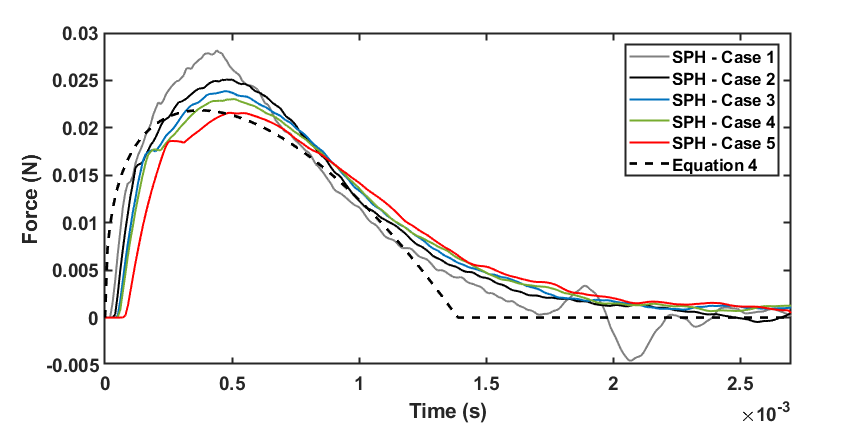}	
\caption{The temporal evolution of the impact force on the surface of the solid material is compared between Case $1$ to Case $5$ as predicted by the SPH model. The analytical solution of Equation \ref{temp1}, which is the impact force under the liquid droplet assuming a rigid surface, is also superimposed.}\label{para_for2}
\end{figure}

Figure \ref{para_for2} presents the total contact force on the surface of the substrates from Case $1$ to Case $5$ obtained by the SPH model, together with the analytical impact force of Equation \ref{temp1}. As the solid materials' Young's modulus decreases, we see a trend of delay in force initiation, followed by decreasing force peaks indicating a mitigating impact intensity as the substrate material becomes softer. Interestingly, the mitigation in impact intensity induced by softer materials makes the benchmark predictions from the SPH model closer to the analytical impact force (Equation \ref{temp1}). The tails of all cases show almost the same impact force histories in time, including the softest case (Case $5$) studied. It is worth-noting that the SPH simulations for softer material cases (here Cases $4$ and $5$) require much finer SPH particles due to the excessive deformations of the very soft substrate by single particle impact, while the ANCM simulations show no problems.

\begin{figure}[t]
\centering
\includegraphics[width=9.5 cm]{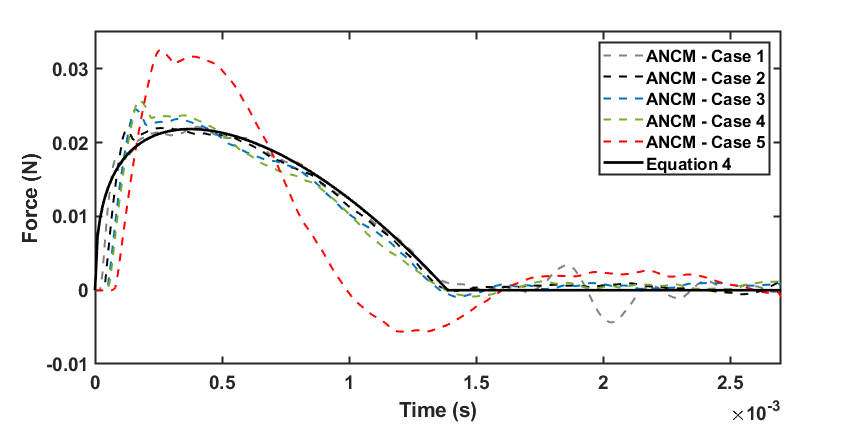}	
\caption{The temporal evolution of the impact force on the surface of the solid material is compared between Case $1$ to Case $5$ as predicted by the ANCM model. The analytical solution of Equation \ref{temp1}, which is the impact force under the liquid droplet assuming a rigid surface, is also superimposed.}\label{para_for}
\end{figure}

We now compare the total contact force on the surface of the substrates for Cases $1$ to $5$ of Table \ref{table_para} obtained by the ANCM models in Figure \ref{para_for}, together with the analytical impact force of Equation \ref{temp1}. Although the same impact loading of the analytical pressure of Equation \ref{pre} was used, the computed contact forces from the FE analysis of respective solid surfaces are not the same. As can be seen, the computations for the reference case (Case $1$) calculates the exact shape of the analytically calculated impact force, which leads to a reasonable deduction that, at Young's modulus of $47,400$ \unit{\pascal}, the deformation at the solid surface has a negligible effect to the droplet impact pressure, and hence justifies the applicability of the rigid surface assumption. Comparing the analytical calculation to the SPH results (Figure \ref{para_for2}), a delay in initiation of force can be observed as a resistance to the impact. However, instead of mitigating impact intensity by decreasing force maxima, there is a slight force increase before the usual force peak time (around $0.5$ \unit{\milli\second}) as the substrate material becomes softer. This increase leads to an emerging force maximum at impact time around $0.16-0.25$ \unit{\milli\second}. These observations are attributed to material compliance for the softer substrates where impacts are accompanied by surfaces motion due to deformation by the impact, causing time delay in force increase (also see Section \ref{3.2.3}). Despite the compliance difference at the beginning of impacts for the Cases $1$ to $4$, the impact force time histories remain almost the same. However, Case $5$ shows a large discrepancy relative to the previous cases in the temporal evolution of the tail shape after the force maximum, that may indicate a fundamental change in model behaviour for $E=4,740$ \unit{\pascal}. To investigate the origin of this behaviour, we next examine the displacements on the material surfaces and throughout the substrate.

\subsubsection{Displacements on the surface and in the substrate}\label{3.2.2}
\begin{figure*}
    \centering
    \begin{subfigure}[b]{0.48\textwidth}
        \centering
        \includegraphics[width=7 cm]{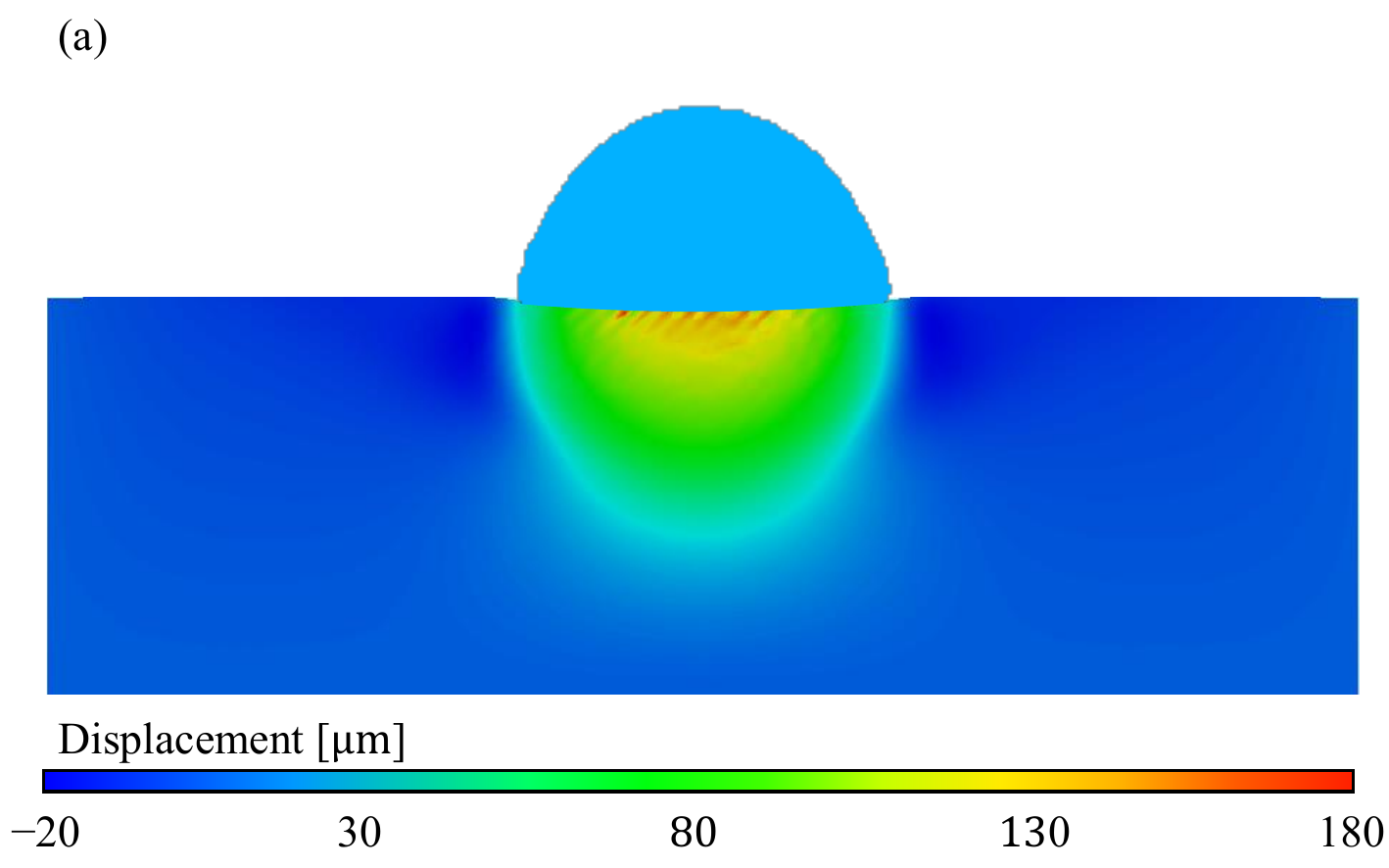}
    \end{subfigure}
    \begin{subfigure}[b]{0.48\textwidth}
        \centering
        \includegraphics[width=7 cm]{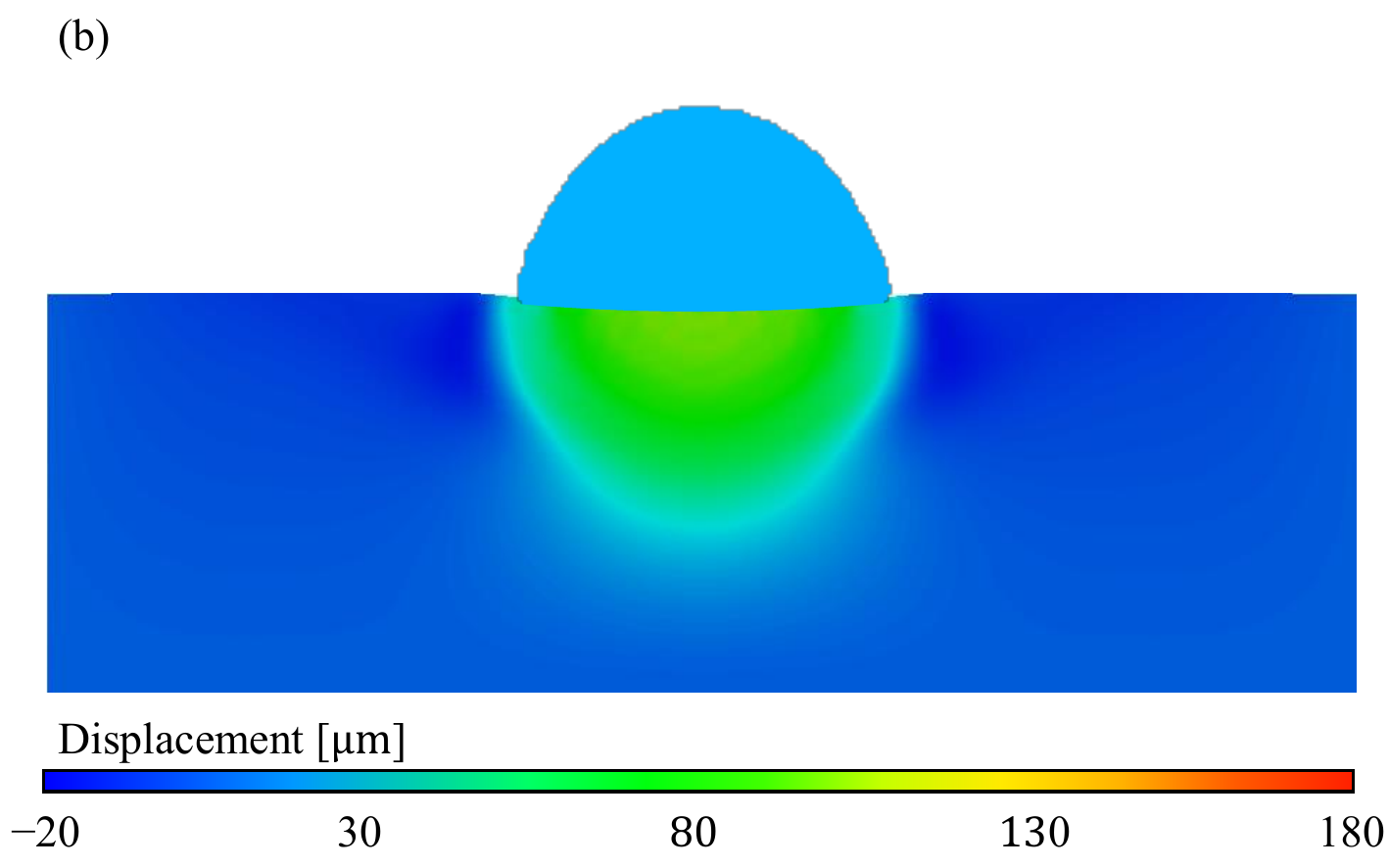}
    \end{subfigure}
    \begin{subfigure}[b]{0.48\textwidth}
        \centering
        \includegraphics[width=7 cm]{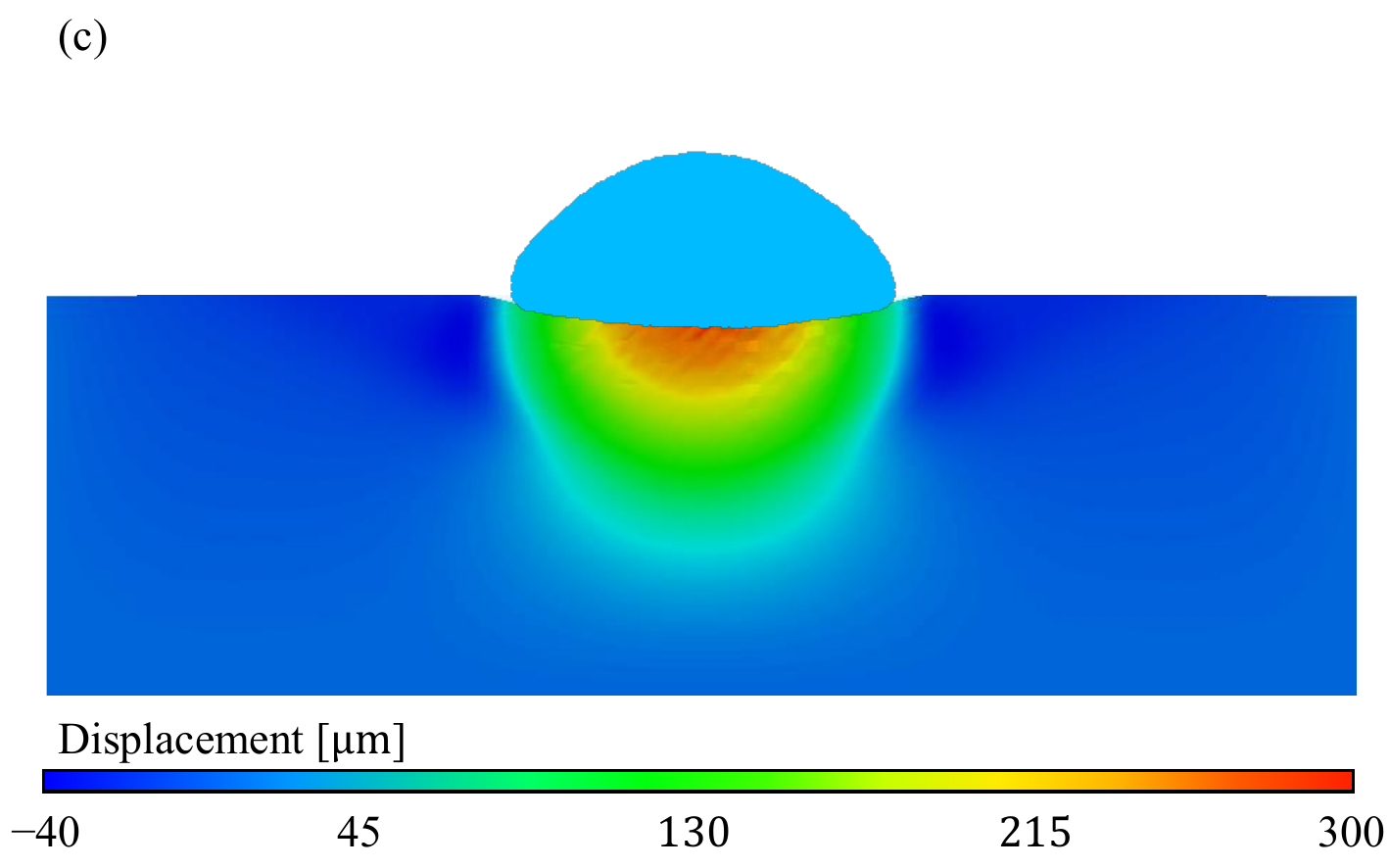}
    \end{subfigure}
    \begin{subfigure}[b]{0.48\textwidth}
        \centering
        \includegraphics[width=7 cm]{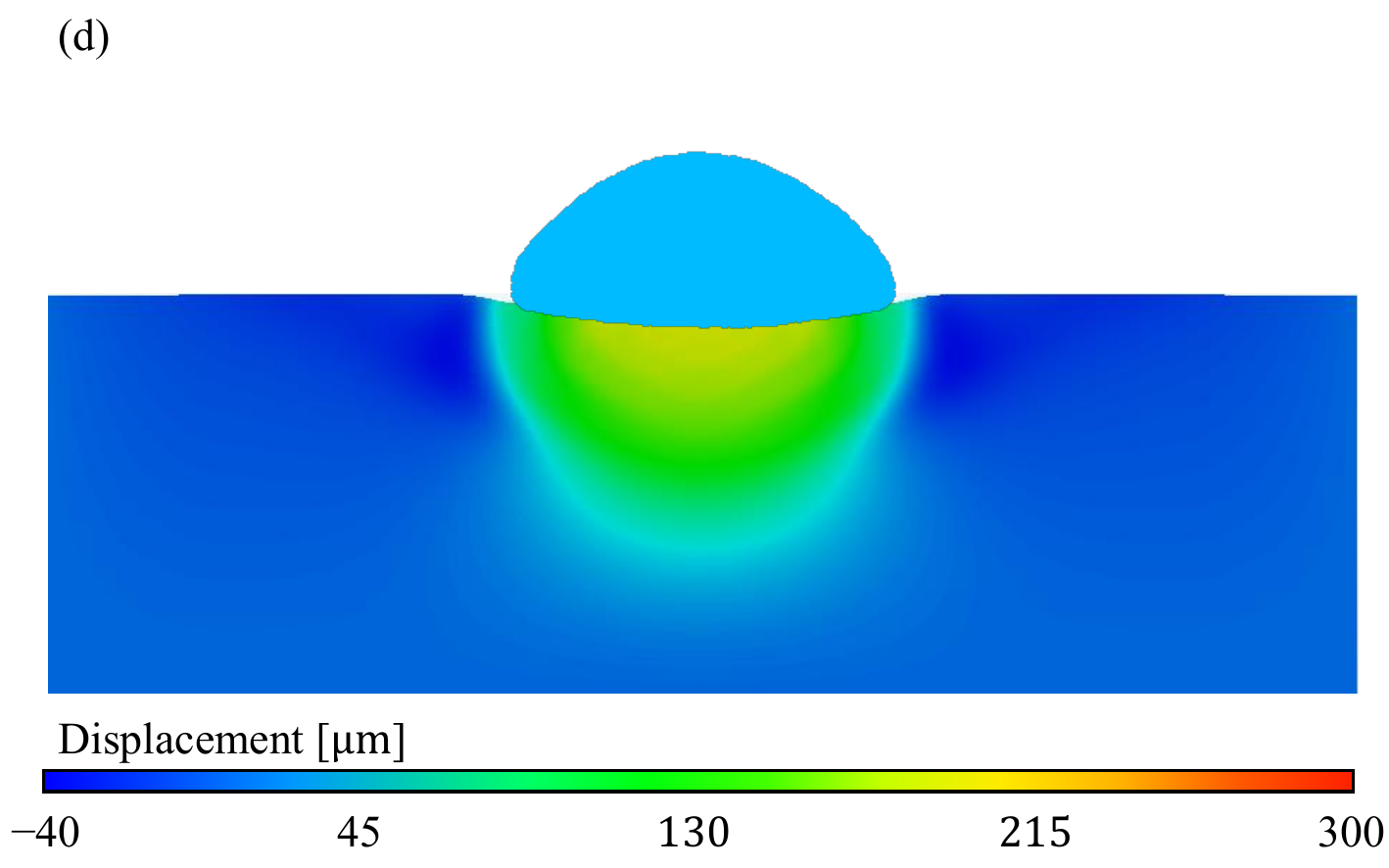}
    \end{subfigure}
    \begin{subfigure}[b]{0.48\textwidth}
        \centering
        \includegraphics[width=7 cm]{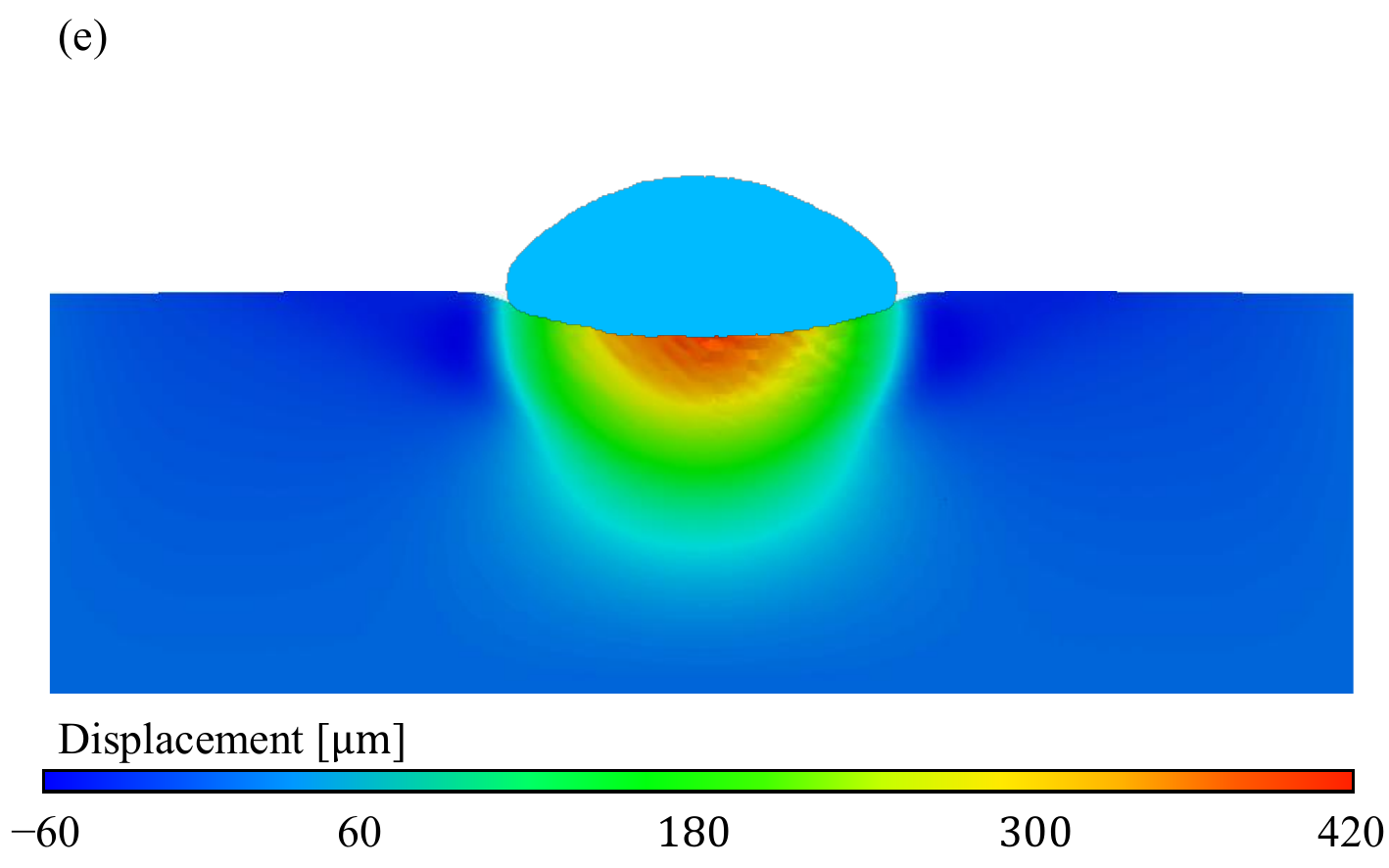}
    \end{subfigure}
    \begin{subfigure}[b]{0.48\textwidth}
        \centering
        \includegraphics[width=7 cm]{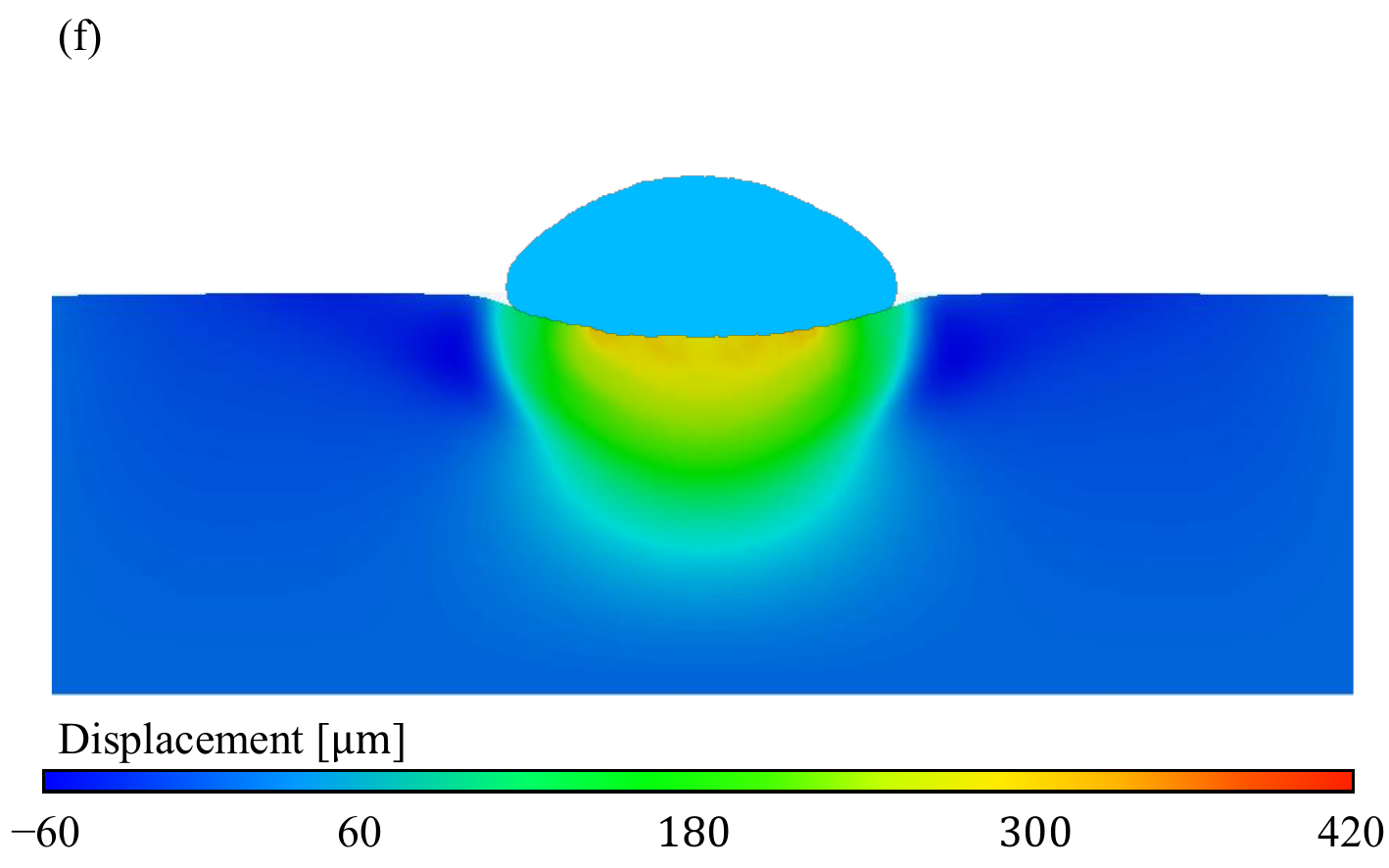}
    \end{subfigure}
    \begin{subfigure}[b]{0.48\textwidth}
        \centering
        \includegraphics[width=7 cm]{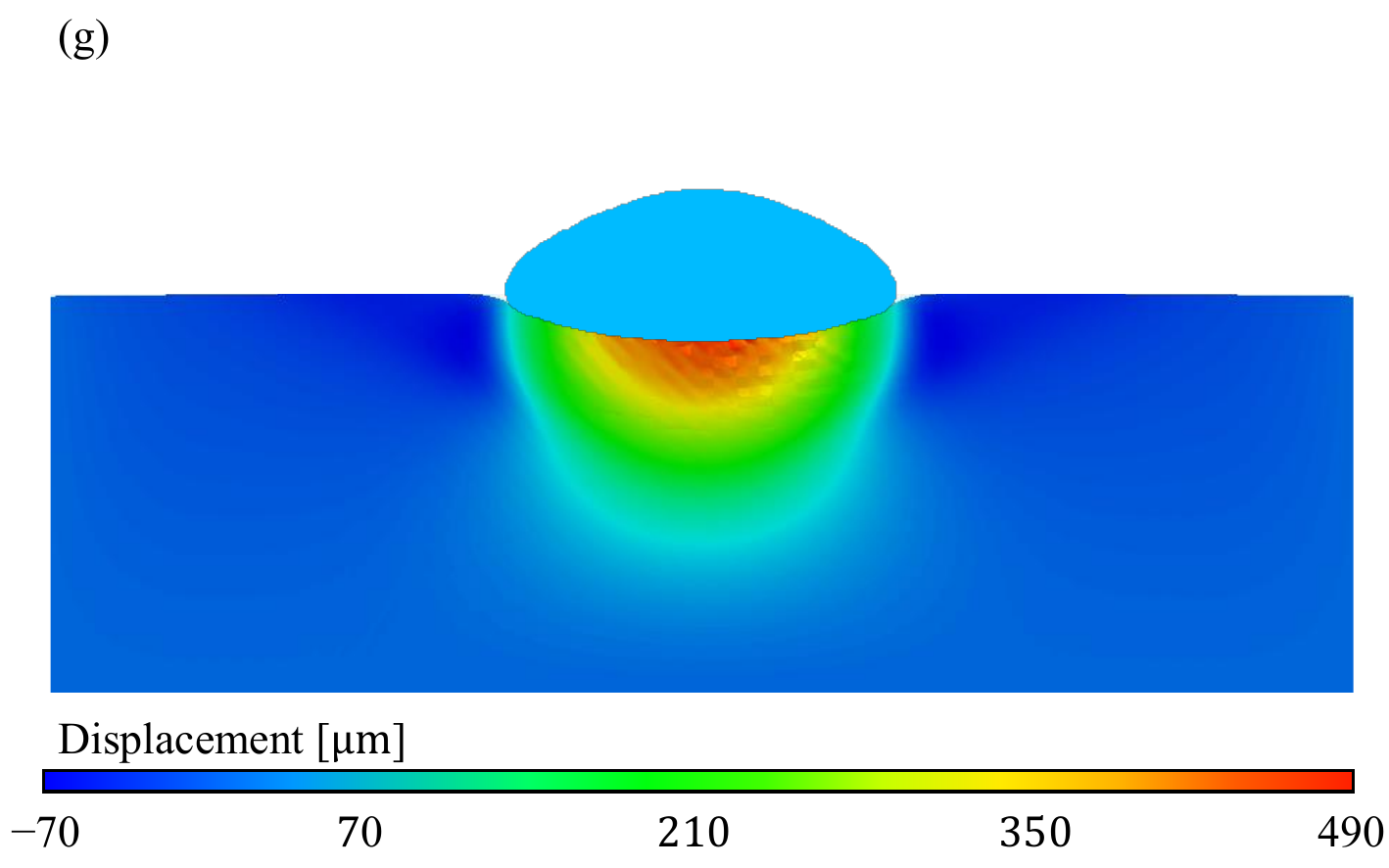}
    \end{subfigure}
    \begin{subfigure}[b]{0.48\textwidth}
        \centering
        \includegraphics[width=7 cm]{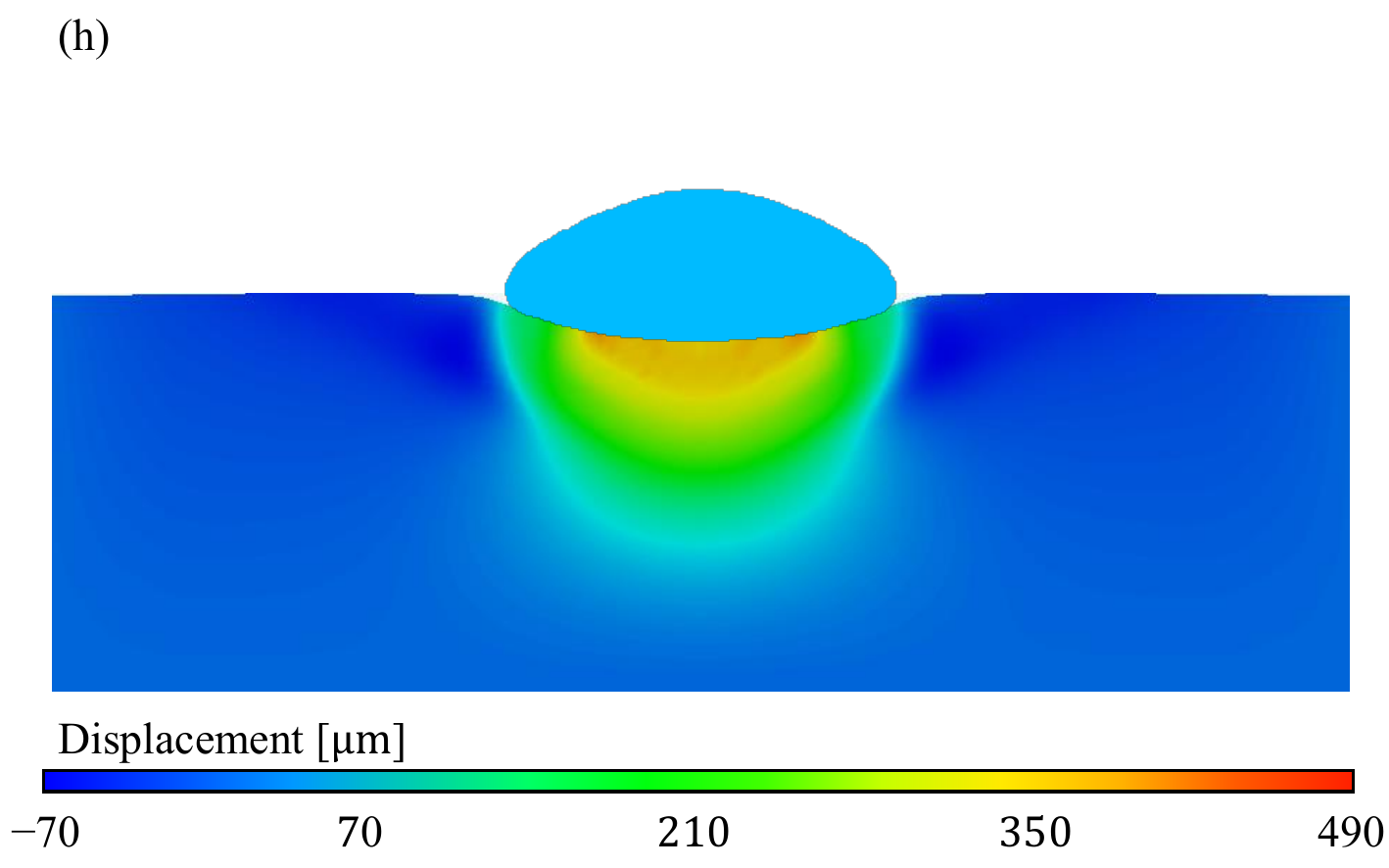}
    \end{subfigure}
    \begin{subfigure}[b]{0.48\textwidth}
        \centering
        \includegraphics[width=7 cm]{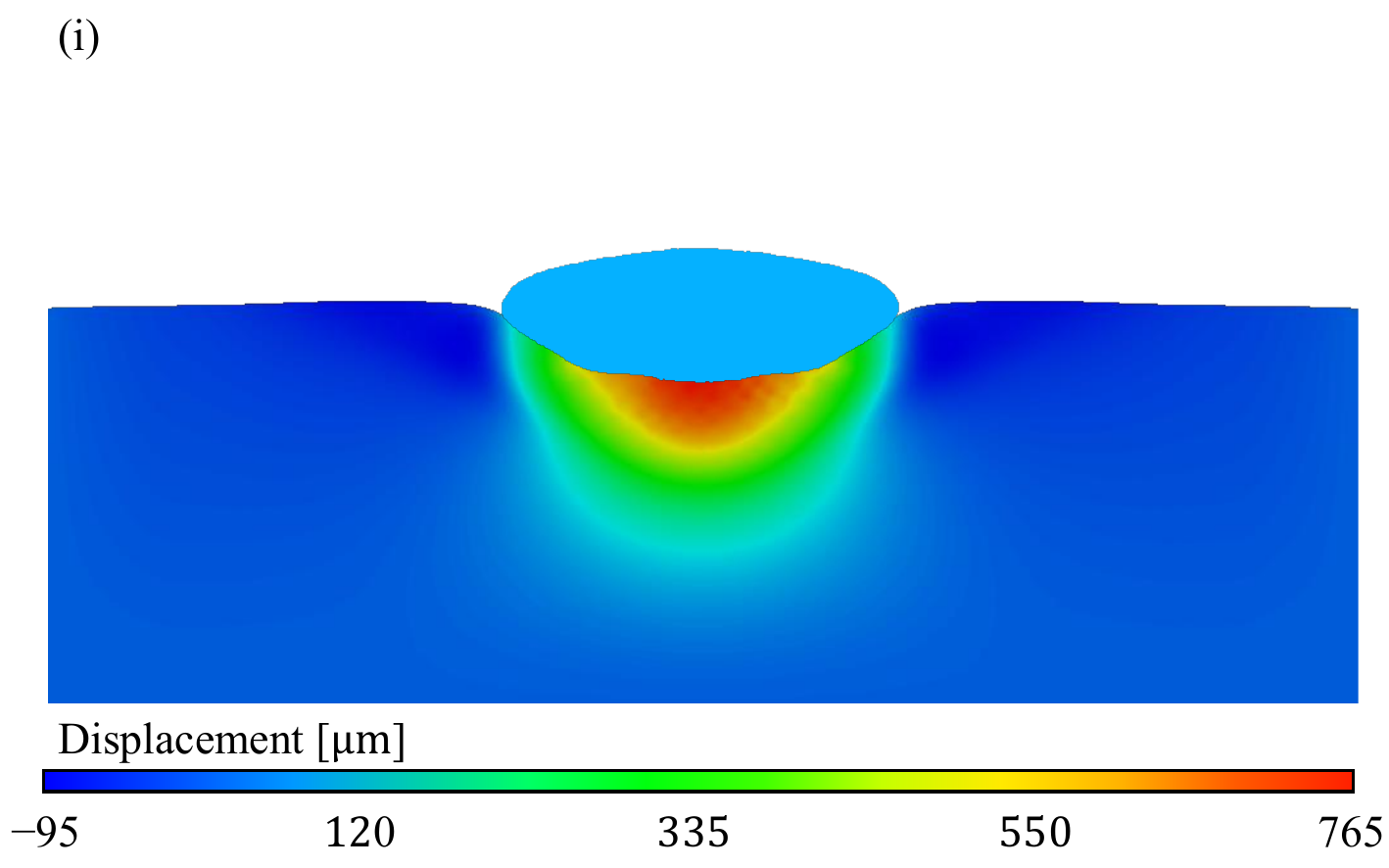}
    \end{subfigure}
    \begin{subfigure}[b]{0.48\textwidth}
        \centering
        \includegraphics[width=7 cm]{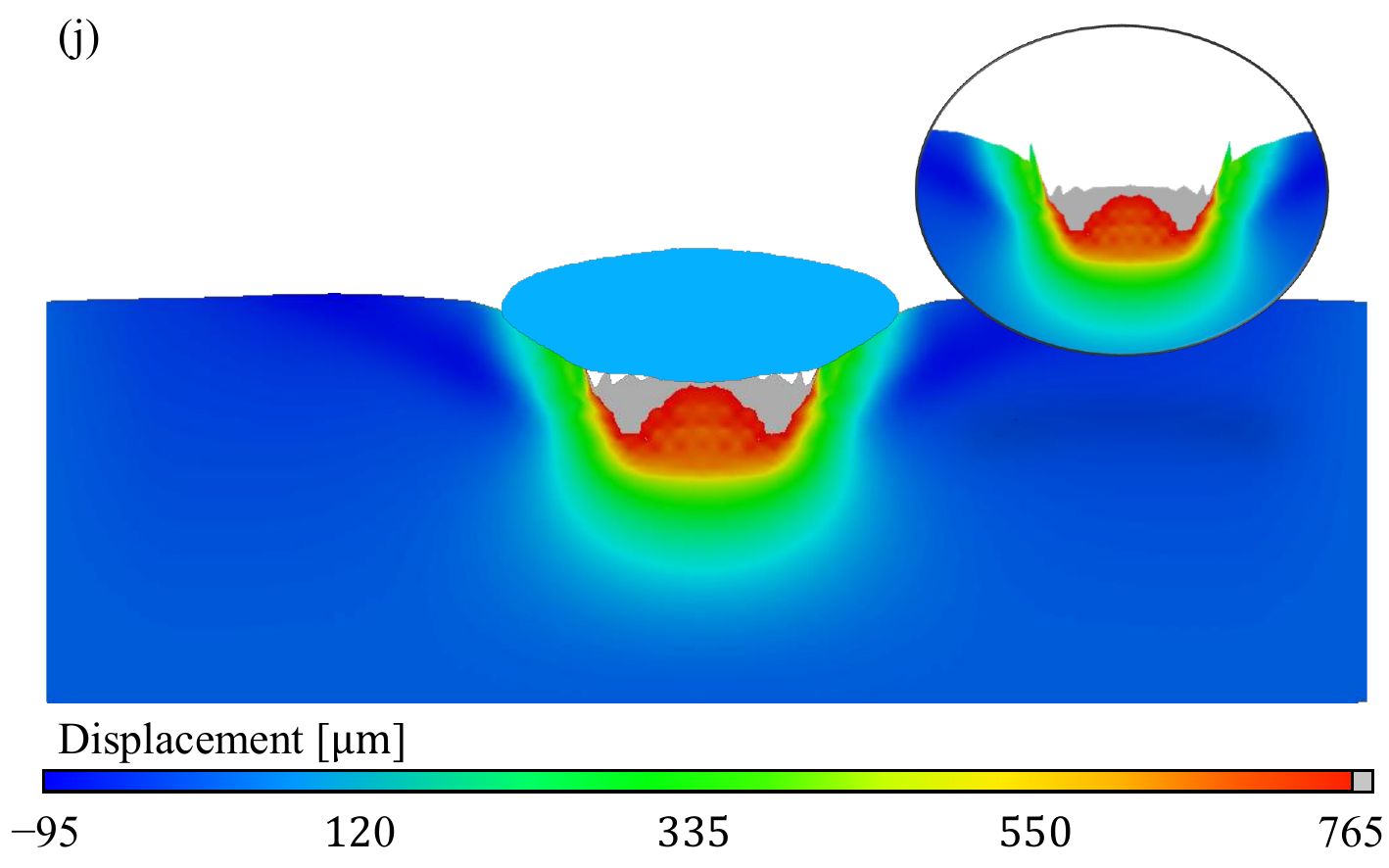}
    \end{subfigure}
    \caption{Spatial distributions of (vertical) displacements within the solid materials obtained from the SPH (left column) and ANCM (right column) models at different material stiffness and different times are compared for Case $1$ (a, b) at $T=590$ \unit{\micro\second}, Case $2$ (c, d) at $T=860$ \unit{\micro\second}, Case $3$ (e, f) at $T=1040$ \unit{\micro\second}, Case $4$ (g, h) at $T=1080$ \unit{\micro\second} and Case $5$ (i, j) at $T=1420$ \unit{\micro\second}. Each subfigure shows the front view with a view cut at the $y$-$z$ plane of symmetry, \emph{i.e.} the same viewing position as in Figure \ref{figure_mesh}. The same colorbar limits have been used for the SPH (left column) and ANCM (right column) results at each impact time. For illustration purpose, droplet configurations are added to the ANCM results from the SPH simulation at corresponding times. Note that the droplet color is independent of the displacement colormap. Inset of subfigure (j) shows the steeper (almost vertical) walls.}
\label{result_para}
\end{figure*}

The previous analysis of the contact force on the solid material surface showed a discrepancy between SPH and ANCM computations regarding the material compliance, and hence surface deformation, at the beginning of the impact. Figure \ref{result_para} presents the vertical displacement of the substrates at different times for Case $1$ (Figure \ref{result_para}a and \ref{result_para}b at $T=590$ \unit{\micro\second}), Case $2$ (Figure \ref{result_para}c and \ref{result_para}d at $T=860$ \unit{\micro\second}), Case $3$ (Figure \ref{result_para}e and \ref{result_para}f at $T=1040$ \unit{\micro\second}), Case $4$ (Figure \ref{result_para}g and \ref{result_para}h at $T=1080$ \unit{\micro\second}) and Case $5$ (Figure \ref{result_para}i and \ref{result_para}j at $T=1010$ \unit{\micro\second}) computed from SPH (left column) and ANCM (right column). The time instants for each case are at maximum vertical displacements. As expected, surfaces deform more for softer substrate material, leading to formation of a crater for Case $5$ where a surface dimple existed for the reference case (Case $1$). From Cases $1$ to $4$, the calculated displacements from the ANCM model agree well with the results of SPH, except a slight overestimation in the SPH model at the center of the surface. This overestimation in surface displacements comes from the gap in surface contact loadings as seen in Figure \ref{force_ref}. Nevertheless, the differences for Cases $1$ to $4$ are very small so that the surface geometries are hardly distinguished. Large difference, however, is observed for Case $5$ between the displacements maxima computed by the two models since the ANCM overpredicts the vertical displacements unexpectedly. In particular, the surface crater for the ANCM computations has steeper (almost vertical) walls (Figure \ref{result_para}j), at the bottom of which, we observe the highest material displacements.

To understand the formation of steep craters where dimples existed, we recall that, in the ANCM model, we coupled the analytical pressure solution \ref{pre} with solid material FE analysis as a time- and spatial-dependent loading on the surface of the FE substrate (Section \ref{2.2}). Specifically, in FE analysis, the analytical loading on each individual surface element is dependent on the element's radial position $r=\sqrt{x^2+y^2}$ at time $t$, specified through the user-defined VDLOAD-subroutine in ABAQUS. Interestingly, this subroutine calculation process does not consider the $z$-coordinate of individual elements as the analytical solution assumes the rigid surface.

\begin{figure}
    \centering
    \begin{subfigure}[b]{0.23\textwidth}
        \includegraphics[width=4.5 cm]{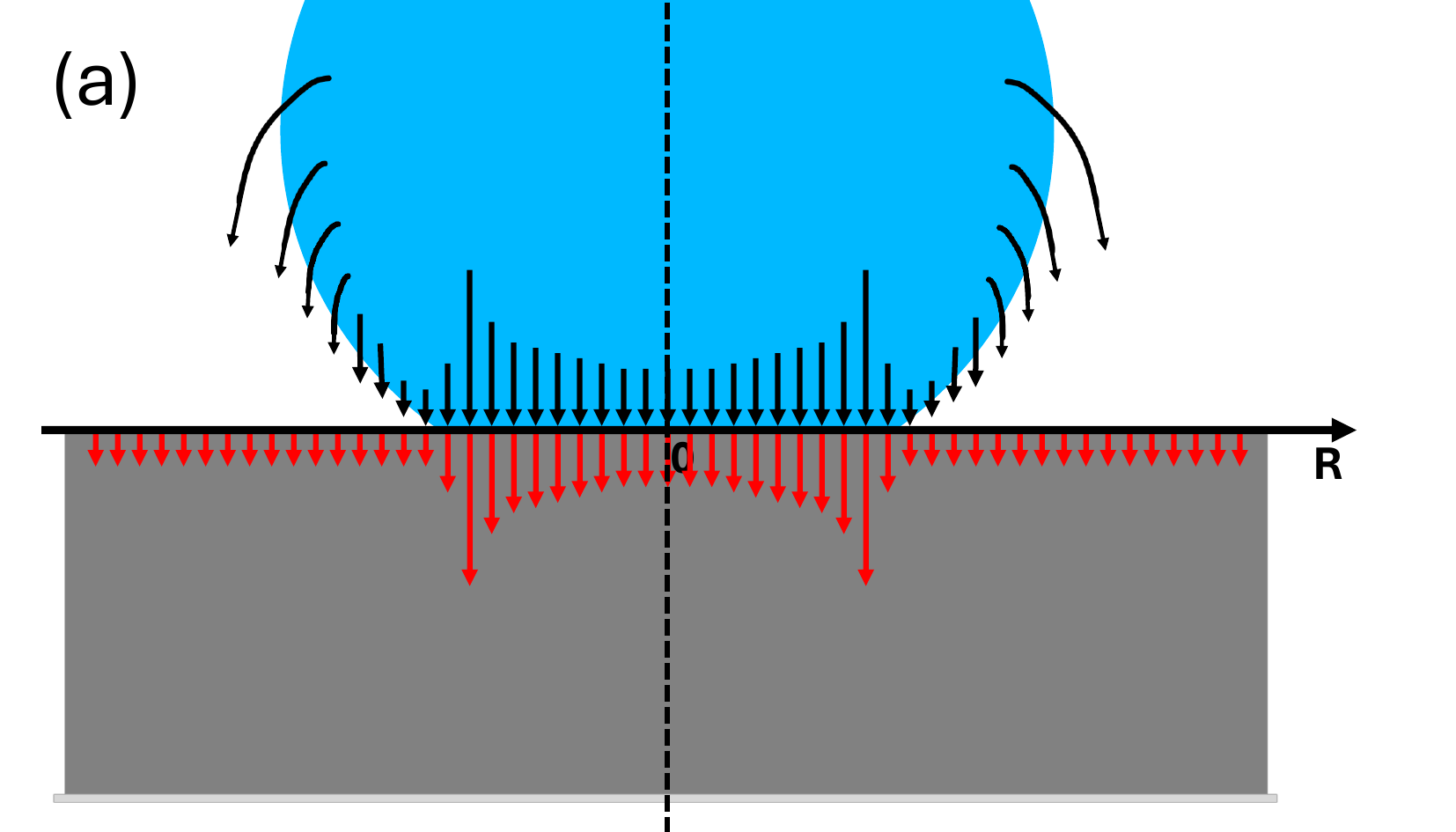}
    \end{subfigure}
    \begin{subfigure}[b]{0.23\textwidth}
        \includegraphics[width=4.5 cm]{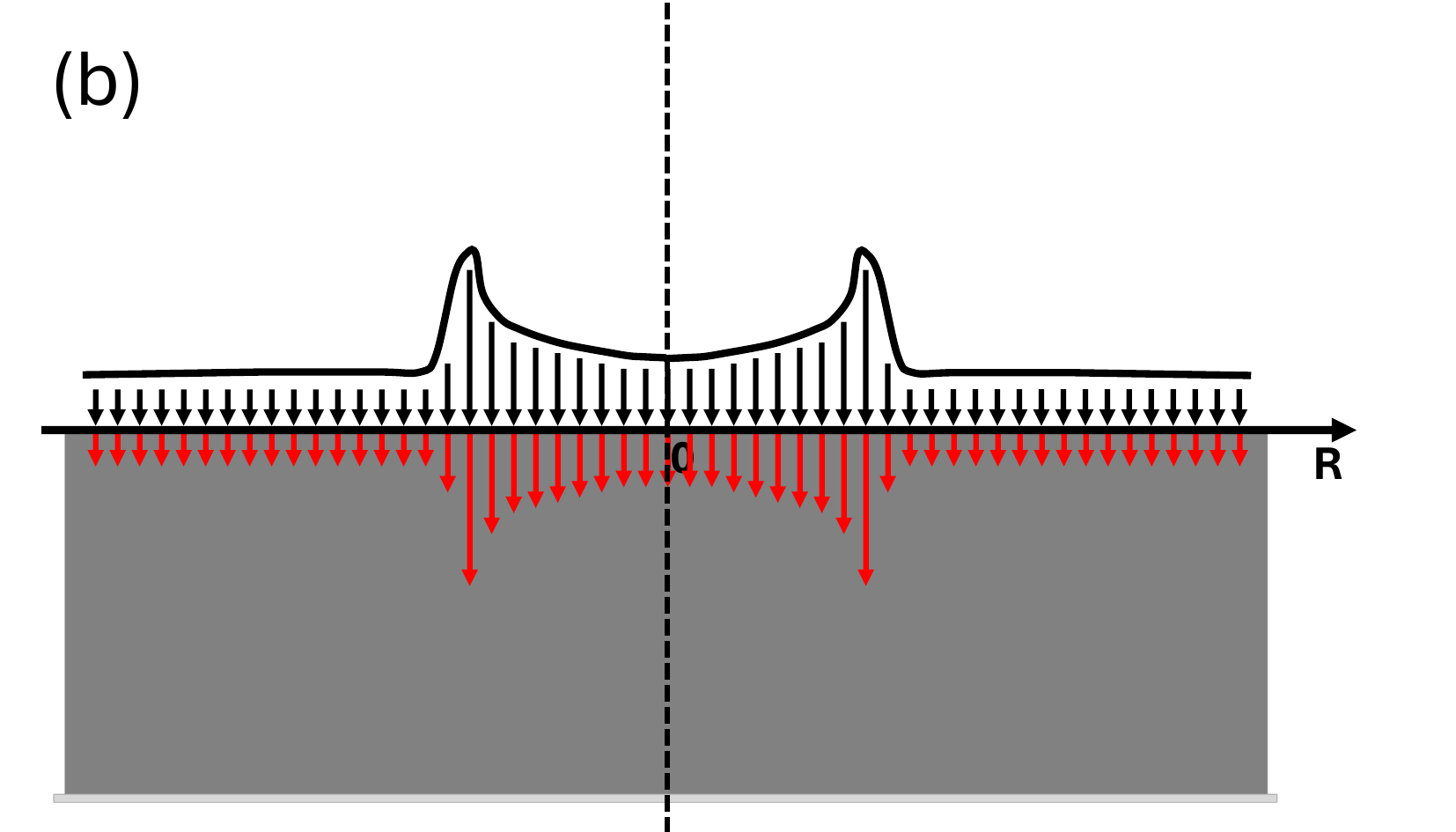}
    \end{subfigure}
    \begin{subfigure}[b]{0.23\textwidth}
        \includegraphics[width=4.5 cm]{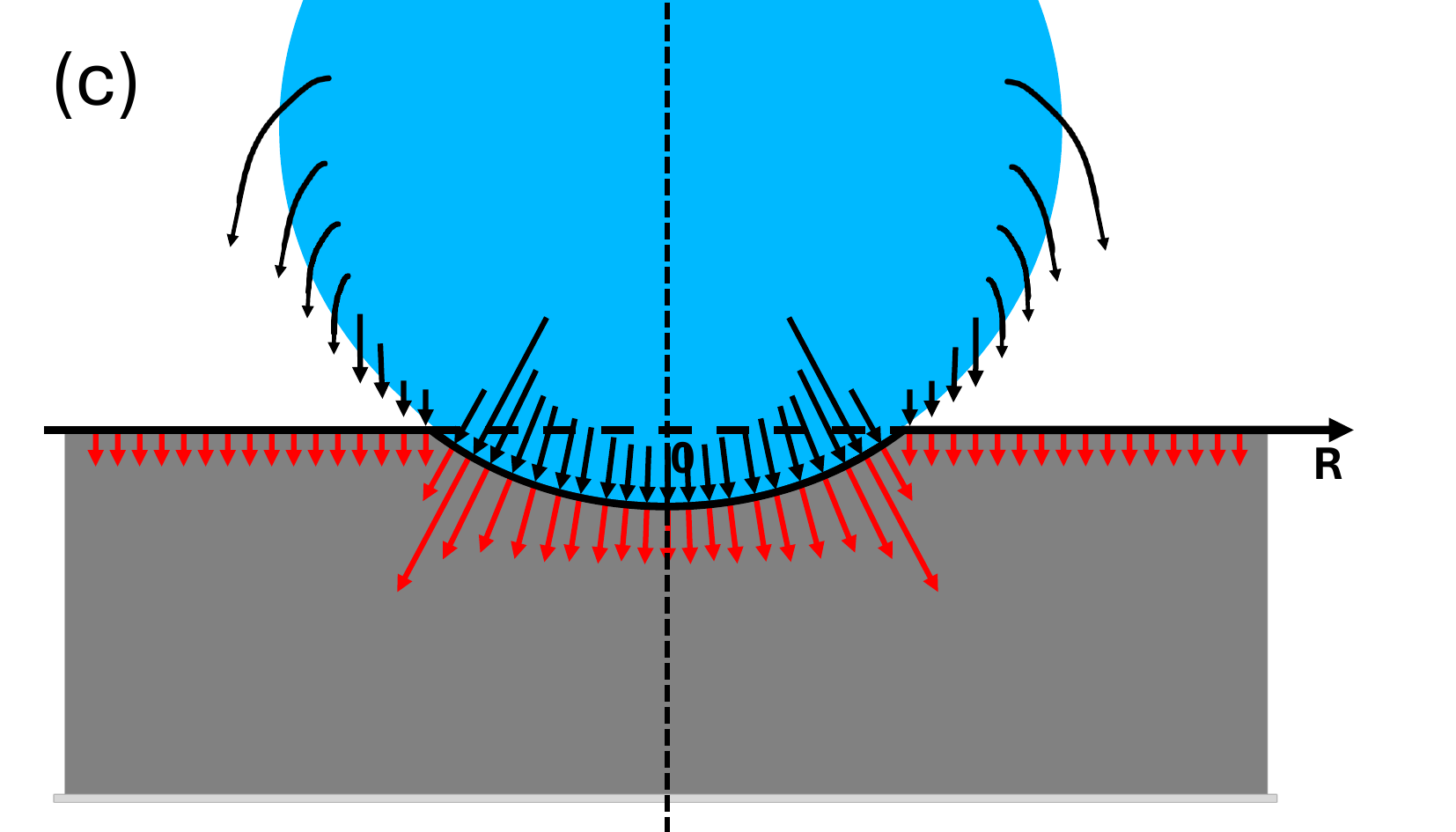}
    \end{subfigure}
    \begin{subfigure}[b]{0.23\textwidth}
        \includegraphics[width=4.5 cm]{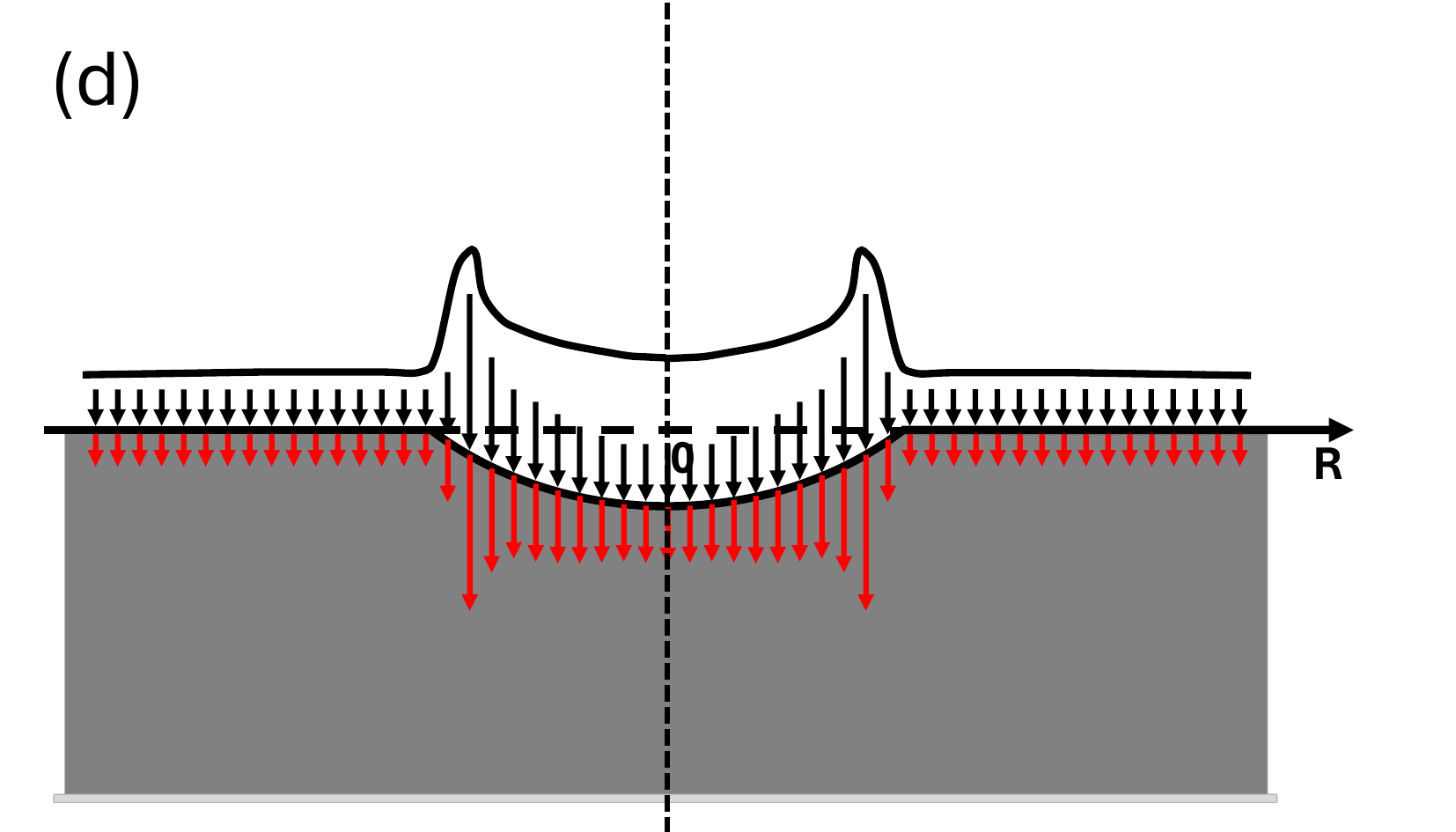}
    \end{subfigure}
    \caption{Illustration of the loading mechanisms of a liquid droplet impact onto the surface of rigid (a, b) and soft (c, d) solid materials in SPH (left column) and ANCM (right column) models. Arrow lengths approximate the magnitudes of impact pressure.}
\label{mechanism}
\end{figure}

Figure \ref{mechanism} presents the numerical loading conditions upon droplet impact in the SPH model, or analytical pressure impact (which models the impact of a liquid droplet) in the ANCM model. In the case of (relatively) rigid surface (Figure \ref{mechanism}a and \ref{mechanism}b), the ANCM model applies the analytical pressure solution of Equation \ref{pre} onto the surface of the substrate as a function of individual element's radial position, as explained above. Since the surface is (relatively) rigid, all elements on the surface have $z$-coordinates equal, or close enough to, zero, which makes the numerical scenario the same, or close enough, to the surface boundary conditions in derivations of the analytical solution~\cite{Hao_ana}. In this way, the analytical pressure solution in the ANCM model (Figure \ref{mechanism}b) describes the intended pressure loadings as if impacted with a liquid droplet in the SPH model (Figure \ref{mechanism}a). This explains the working mechanism for relatively rigid surface (with criterion provided below).

For (relatively) soft surface (Figure \ref{mechanism}c and \ref{mechanism}d), the liquid droplet in the SPH model (Figure \ref{mechanism}c) explores craters on the surface through fluid dynamics of impacting, spreading and splashing. The fluid dynamics is sensitive to the surface geometry and vice versa. Hence the two-way coupling algorithms in soft material impact simulations are more significant than ever. Specifically, on the surface of the craters, which is also a part of the impact surface, impact forces are diverted and spread by the curved geometry (black arrows in Figure \ref{mechanism}c) in normal directions to the crater surface. As a result, the total contact force on impact surface in vertical direction is greatly mitigated. This explains the decreasing force peaks that we found in Section \ref{TCF} (Figure \ref{para_for2}) as substrate materials become softer. 

While for the ANCM model (Figure \ref{mechanism}d), as explained before, the user-defined VDLOAD-subroutine does not consider the spontaneous $z$-coordinate of each element on the impact surface, therefore the exact analytical pressure of Equation \ref{pre} is applied in the same way for all substrate geometries. As a result, integrating the same loaded impact pressure over substrate surface area and time, we obtain the same impact impulse, or momentum change, applied to the material substrate. In other words, the total area under the force-time curve, which is impact impulse, is conserved. This explains the emerging force spikes obtained by the ANCM computations at impact times around $0.16-0.25$ \unit{\milli\second} in Figure \ref{para_for} as substrate material becomes softer\textemdash since the increasing material compliance delays the resistance force while impact impulse has to be conserved. Same impulse conservation is the reason behind the fundamentally different tail of Case $5$ in Figure \ref{para_for} as the force peak has been overshot significantly. 

Finally, the reason behind the fundamental changes of surface deformation at (relatively) soft material is explained. On the surface of the craters, analytical impact pressure is exerted vertically downwards (black arrows in Figure \ref{mechanism}d) as if on the rigid surface (Figure \ref{mechanism}b) in the ANCM model. This leads to a non-physical amount of vertical loads on the substrate, comparing the red arrows in Figure \ref{mechanism}c and \ref{mechanism}d. In particular, at the rim of the crater, impact loads are the highest. This leads to the steep-wall craters (Figure\ref{result_para}j) observed in ANCM computation comparing to the inclined-wall crater in SPH computation (Figure\ref{result_para}i) where impact loads at the rim are deviated into horizontal components. This further explains the fundamentally different overshoot of vertical total contact force maxima (Figure \ref{para_for}), and a change-over of deformation maxima (see below), between the SPH and ANCM computations.

\begin{figure*}[t]
\centering
\includegraphics[width=16 cm]{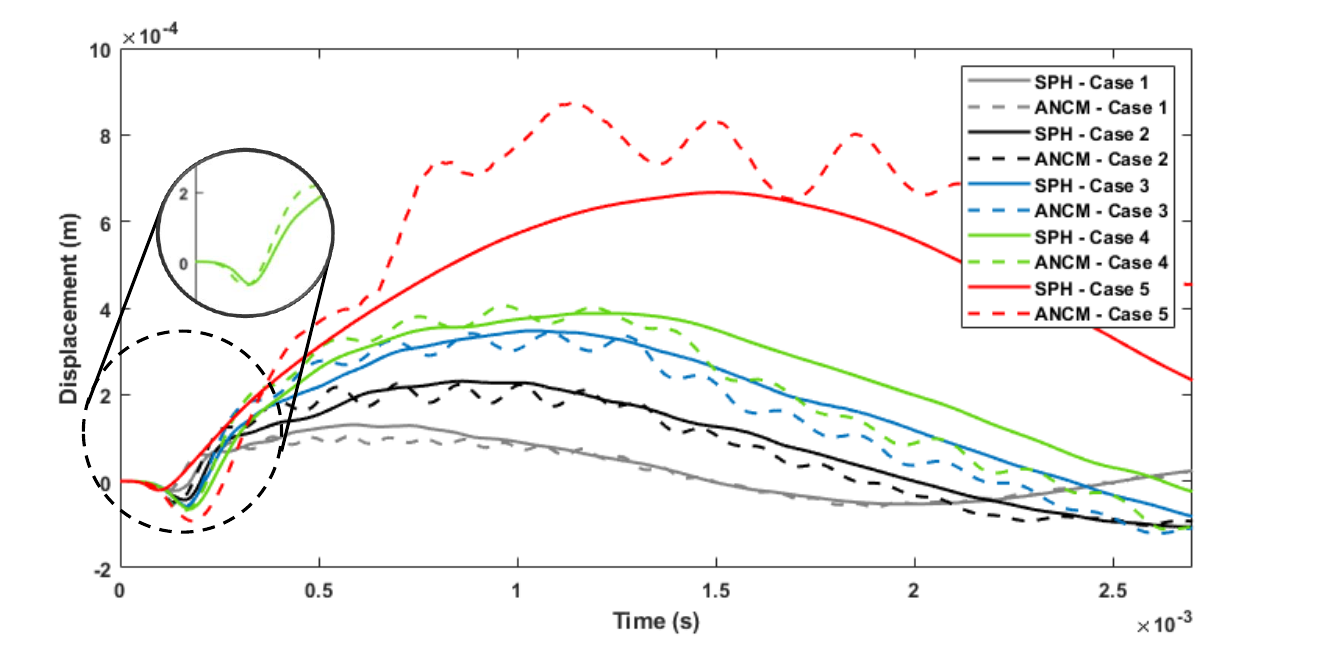}	
\caption{Temporal evolution of the vertical displacements at radial position $R=1.0$ \unit{\milli\meter} on the surface of the solid material for the five cases of Table \ref{table_para} which have different material stiffness computed by SPH (solid lines) and ANCM (dashed lines) models. The inset isolates Case $4$ at the start of the impact.}\label{para_U}
\end{figure*}

\subsubsection{Applicability limit of the ANCM model}\label{3.2.3}
Based on the mechanism proposed in Section \ref{3.2.2} (Figure \ref{mechanism}), we now determine the applicability limit of the ANCM model with insights from material surface deformation. Figure \ref{para_U} presents the vertical displacement histories at radial positions of two-thirds ($1.0$ \unit{\milli\meter}) of initial droplet radius for each case. The radial position is selected based on the displacement results of Section \ref{3.1}, where we observed the highest surface deformation deflections at radial positions between one-third ($0.5$ \unit{\milli\meter}) and two-thirds ($1.0$ \unit{\milli\meter}) of initial droplet radius in a ring-shape pattern. As can be seen from Figure \ref{para_U}, for Cases $1$ to $4$, displacements obtained by ANCM are in good agreement with the benchmark displacements obtained by SPH. Surface elevation (\emph{i.e.}, negative displacement) and depression (\emph{i.e.}, positive displacement) both increase in magnitudes as the substrate material becomes softer. However, Case $5$ exhibits a pronounced overshoot in surface depression, which can be traced back to a less apparent but equally important overshoot in the initial elevation. For this reason, we use the initial surface elevation as a representative measure of the subsequent deformation behaviour. 

This brings our attention to the specific Case $4$, focused in the inset of Figure \ref{para_U}, where the initial surface elevation obtained by ANCM protrudes almost the same amount as the results from SPH. We regard this material softness level (Case $4$ of $E=10,000$ \unit{\pascal}) as the critical limit of the ANCM model. Above this limit, stiffer materials with Young's modulus greater than $10,000$ \unit{\pascal} exhibit surface depression of a physically reasonable magnitude in the ANCM model (see gray lines in Figure \ref{para_U}), consistent with the analytical solution (Figure \ref{force_ref}). Thus, within this range, the model is safe to apply with the reasonable assumption of rigid surface. For softer materials with Young's modulus below $10,000$ \unit{\pascal}, the ANCM model begins to exhibit rapid over-depression relative to the SPH benchmark (see red lines in Figure \ref{para_U}), along with fundamentally different deformation modes, including the formation of steep-walled craters (Figure \ref{result_para}j). 

It should be noted, however, that the rapid change beyond the softness limit is not sudden. The limit value $E=10,000$ \unit{\pascal} is chosen, out of conservative consideration, as the meeting point between decreasing impact intensity of SPH and increasing impact intensity of ANCM as material becomes softer. Therefore, with care, the ANCM model may still be applicable for materials with stiffness lower, but closer, to $E=10,000$ \unit{\pascal}, subject to the applications and impact conditions.

\section{Conclusion}
\label{conclusion}
In this paper, the developed ANCM model~\cite{Hao_ANCM} for simulating liquid droplet impacts on solid materials has been validated and studied specifically on soft materials. The purpose of the study is to find the applicability limit of the developed model on soft materials with surface deformation as in engineering applications, instead of the ideal rigid surface. Key findings are summarized below:
\begin{itemize}
    \item The developed ANCM model is validated against experimental data on a urethane gel phantom materials at Young's modulus of $47,400$ \unit{\pascal}. Comparison between the simulation results and the analytical impact force shows that, for materials at Young's modulus of $47,400$ \unit{\pascal} or stiffer, the developed ANCM model works to the expectation as on an assumed rigid surface.
    \item As the substrate becomes softer, there is a delay in contact force initiation due to material compliance. Following the force initiation, the SPH model predicts a decreasing contact force maxima due to higher material compliance, while the ANCM model predicts a slight force accumulation in time due to total impact impulse conservation.
    \item The difference between the SPH and ANCM models as substrate becomes softer is caused by the ways impact loads are exerted on non-flat surface. In the SPH, impacting liquids exert loads normal to the contact surface, and hence alleviate the total impact force towards radial direction. While the analytical loads in ANCM are exerted independently of the surface geometries, providing the same amount of impact impulse at each impact.
    \item The different impact mechanisms of the ANCM model to the fully-numerical SPH model leads to a fundamental breakdown on substrate material at Young's modulus $E=4,740$ \unit{\pascal} (the softness of gelatine~\cite{Yuto2023}), as a result of non-physical overshoot in impact intensity including impact force and surface displacements. We observe a switch in surface displacement maxima between the two numerical models at $E=10,000$ \unit{\pascal}. Therefore, we regard $E=10,000$ \unit{\pascal} as the onset of the ANCM model failure zone, below which threshold, excessive total contact forces, non-physical surface deformations and steep wall craters happen.
\end{itemize}

\noindent The future work of the model:
\begin{itemize}
    \item In current ANCM model, VDLOAD subroutine in ABAQUS specifies analytical impact pressure independent of surface element's height or depth. This leads to the invariance of impact loads to the surface deformation of soft materials, and hence the failure of the ANCM model on soft material substrates. Future work shall modify the VDLOAD subroutine to three dimension to consider the surface element's height or depth, thereby to apply impact loads normal to the contact surface as if in real impacts. This could potentially resolve the applicability limits of the ANCM model on very soft materials.
\end{itemize}

\section*{Author Contributions}
Data curation, H.H.; conceptualization, H.H. Y.H.; formal analysis, H.H. A.S. A.T. Y.H. M.C.; funding acquisition, H.H. Y.H.; investigation, H.H. A.S. A.T. Y.H. M.C.; methodology, H.H. A.S. A.T. Y.H. M.C.; project administration, Y.H.; resources, H.H. A.S. A.T. Y.H. M.C.; software, H.H. A.S. A.T. Y.H. M.C.; supervision, A.S. A.T. Y.H. M.C.; validation, H.H.; visualization, H.H.; writing - original draft, H.H.; writing - review and editing, H.H. Y.H. M.C..

\section*{Declaration of competing interest}
The authors declare that they have no known competing financial interests or personal relationships that could have appeared to influence the work reported in this paper.

\section*{Funding}
H.H. was funded by the Mechanical Engineering PhD Scholarship from the Department of Mechanical Engineering of Imperial College London.

\section*{Acknowledgments}
The authors thank Dr. Yokoyama and Professor Tagawa at Tokyo University of Agriculture and Technology for providing relevant experimental data.




\begin{thebibliography}{00}


\bibitem{Pergamalis}
Pergamalis, H.. Droplet impingement onto quiescent and moving liquid surfaces. PhD Thesis, Imperial College London, 2002.
\bibitem{Yu}
Zhang, Y., Dong, Z., Li, C., Du, H., Fang, N. X., Wu, L. and Song, Y.. Continuous 3D printing from one single droplet. Nat. Commun., 2020, 11, 4685.
\bibitem{Panao}
Panao, M. and Moreira, A. L. N.. Experimental study of the flow regimes resulting from the impact of an intermittent gasoline spray. Exp. Fluids., 2004, 37, 834-855.
\bibitem{Bergeron}
Bergeron, V., Bonn, D., Martin, J. Y. and Vovelle, L.. Controlling droplet deposition with polymer additives. Nature, 2000, 405, 772-775.
\bibitem{Gohardani}
Gohardani, O.. Impact of erosion testing aspects on current and future flight conditions. Prog. Aerosp. Sci., 2011, 47, 280-303.
\bibitem{Hao}
Hao, H., Domenech, L. and S{\'a}nchez, F.. Modelling rain erosion surface damage initiation in turbine blades based on inspection data at wind farms. Results Eng., 2026, 29, 109517.
\bibitem{Hao2}
S{\'a}nchez, F., Hao, H., Domenech, L., Hardalupas, Y., Dyer, K., Garcia, V., Charalambides, M. N., Ibanez-Arnal, M., Sergis, A. and Taylor, A. M. K. P.. A review and assessment to the rain erosion damage initiation of wind turbine baldes leading edge protection systems based on laboratory testing data and industrial recommended practice {DNVGL-RP-0573}. Submitted to \textit{Results Eng.}, 2025. DOI: 10.2139/ssrn.5165106.
\bibitem{Tagawa}
Tagawa, Y., Oudalov, N., Ghalbzouri, A. E., Sun, C and Lohse, D.. Needle-free injection into skin and soft matter with highly focused microjets. Lab Chip, 2013, 13, 1357-63.
\bibitem{Shojima}
Shojima, M., Oshima, M., Takagi, K., Torii, R., Hayakawa, M., Katada, K., Morita, A. and Kirino, T.. Magnitude and role of wall shear stress on cerebral aneurysm: Computatonal fluid dynamic study of 20 middle cerebral artery aneurysms. Stroke, 2004, 35, 2500-5.
\bibitem{Field2}
Field, J.. Stress waves, deformation and fracture caused by liquid impact. Phil. Trans. R. Soc. Lond. A, 1966, 260, 86-93.
\bibitem{Engel1}
Engel, O. G.. Waterdrop collisions with solid surfaces. J. Res. Natl. Bur. Stand., 1955, 54, 291-298.
\bibitem{Brunton1}
Bruntuon, J. H. and Hancox, N. L.. The erosion of solids by the repeated impact of liquid drops. Phil. Trans. R. Soc. Lond. A, 1966, 260, 121-139.
\bibitem{Sun}
Sun, T., Alvarez-Novoa, F., Andrade, K., Gutierrez, P., Gordillo, L. and Cheng, X.. Stress distribution and surface shock wave of drop impact. Nat. Commun., 2022, 13, 1703.
\bibitem{Yuto2023}
Yokoyama, Y., Mitchell, B., Nassiri, A., Kinsey, B., Korkolis, Y., Tagawa, Y.. Integrated photoelasticity in a soft material: Phase retardation, azimuthal angle, and stress-optic coefficient. Opt. Lasers Eng., 2023, 161, 107335.
\bibitem{Yuto2024}
Yokoyama, Y., Ichihara, S., Tagawa, Y.. High-speed photoelastic tomography for axisymmetric stress fields in a soft material: Temporal evolution of all stress components. Opt. Lasers Eng., 2024, 178, 108224.
\bibitem{Hrennikoff}
Hrennikoff, A.. Solution of problems of elasticity by the framework method. J. Appl. Mech., 1941, 8, 169-175.
\bibitem{Courant}
Courant, R.. Variational methods for the solution of problems of equilibrium and vibrations. Bull. Am. Math. Soc., 1943, 49, 1-23.
\bibitem{Zhou1}
Zhou, J., Liu, J., Zhang, X., Yan, Y., Jiang, L., Mohagheghian, I., Dear, J. P. and Charalambides, M. N.. Experimental and numerical investigation of high velocity soft impact loading on aircraft materials. Aerosp. Sci. Technol., 2019, 90, 44-58.
\bibitem{XJ}
Zhang, R., Zhang, B., Lv, Q., Li, J., Guo, P.. Effects of droplet shape on impact force of low-speed droplets colliding with solid surface. Exp. Fluids, 2019, 60-64.
\bibitem{Samaras}
Samaras, G., Bikos, D., Skamniotis, C., Cann, P., Masen, M., Hardalupas, Y., Vieira, J., Hartmann, C. and Charalambides, M.. Experimental and computational models for simulating the oral breakdown of food due to the interaction with molar teeth during the first bite. Extreme Mech. Lett., 2023, 62, 102047.
\bibitem{Bikos}
Bikos, D., Samaras, G., Charalambides, M., Cann, P., Masen, M., Hartmann, C., Vieira, J., Sergis, A. and Hardalupas, Y. A micromechanical based finite element model approach to accurately predict the effective thermal properties of micro-aerated chocolate. Innov. Food Sci. Emerg. Technol., 2023, 83, 103227.
\bibitem{Keegan1}
Keegan, M. H., Nash, D. H. and Stack, M. M.. Modelling Rain Drop Impact of Offshore Wind Turbine Blades. Am. Soc. Mech. Eng., 2012, 6, 887-898.
\bibitem{Leon}
Doagou-Rad, S., Mishnaevsky, L. Jr.. Rain erosion of wind turbine blades: computational analysis of parameters controlling the surface degradation. Meccanica, 2020, 55, 725-743. 
\bibitem{Verma1}
Verma, A. S., Castro, S. G. P., Jiang, Z. and Teuwen, J. J. E.. Numerical investigation of rain droplet impact on offshore wind turbine blades under different rainfall conditions: A parametric study. Compos. Struct., 2020, 241, 112096.
\bibitem{Zhou2}
Zhou, Q., Li, N., Chen, X., Xu, T., Hui, S. and Zhang, D.. Analysis of water drop erosion on turbine blades based on a nonlinear liquid-solid impact model. Int. J. Impact Eng., 2009, 36, 1156-1172.
\bibitem{Amirzadeh1}
Amirzadeh, B., Louhghalam, A., Raessi, M. and Tootkaboni, M.. A computational framework for the analysis of rain-induced erosion in wind turbine blades, part I: Stochastic rain texture model and drop impact simulations. J. Wind Eng. Ind. Aerodyn., 2017, 163, 33-43.
\bibitem{Nick2023}
Hoksbergen, T., Akkerman, R., Baran, I.. Rain droplet impact stress analysis for leading edge protection coating systems for wnid turbine blades. Renew. Energy, 2023, 218, 119328.
\bibitem{Hao_ANCM}
Hao., H., Charalambides, M. N., Hardalupas, Y., Sergis, A. and Taylor, A. M. K. P.. An analytical-numerical coupled model of liquid droplet impact on solid material surfaces. Unpublished, 2026. arXiv:submit/7327107.
\bibitem{Hao_ana}
Hao., H., Charalambides, M. N., Hardalupas, Y., Sergis, A. and Taylor, A. M. K. P.. Analytical solution of droplet impact on solid surfaces at short and intermediate times. Accepted by \textit{J. Fluid Mech.}, 2026. DOI: 10.1017/jfm.2026.11320.
\bibitem{Ye}
Ye, J., Zhou, H. and He, K.. A generalized framework of two-way coupled numerical model for fluid-structure-seabed interaction (FSSI): Explicit algorithm. Eng. Geol., 2024, 340, 107679.
\bibitem{Richter}
Richter, T.. Numerical methods for fluid–structure interaction problems. PhD Thesis, University of Heidelberg, 2010.
\bibitem{Ahamed}
Ahamed, M., Atique, S., Munshi, M. and koiranen, T.. A concise description of one way and two way coupling methods for fluid-structure interaction problems. Am. J. Eng. Res., 2017, 6, 86-89.
\bibitem{Wagner}
Wagner, H.. Uber stoß- und gleitvorgange an der oberflache von flussigkeiten. J. Appl. Math. Mech., 1932, 12, 193–215.
\bibitem{Zhang}
Zhang, B., Li, J., Guo, P., Lv, Q.. Experimental studies on the effect of Reynolds and Weber numbers on the impact forces of low-speed droplets colliding with a solid surface. Exp. Fluids, 2017, 58, 125.
\bibitem{Miller}
Miller, G., Pursey, H.. On the partition of energy between elastic waves in a semi-infinite solid. Proc. R. Soc. Lond. A, 1955, 233, 55-59.
\bibitem{Kim}
Kim, H., Kim, J., Kang, H., Kim,  S.. Stress wave propagation in a coated elastic half-space due to water drop impact. J. Appl. Mech., 2000, 68 (2), 346.

\end{thebibliography}
\end{document}